\documentclass{article}

\newif\ifshowchanges
\showchangesfalse

\usepackage{PRIMEarxiv}

\usepackage[utf8]{inputenc} % allow utf-8 input
\usepackage[T1]{fontenc}    % use 8-bit T1 fonts
\usepackage{hyperref}       % hyperlinks
\usepackage{url}            % simple URL typesetting
\usepackage{booktabs}       % professional-quality tables
\usepackage{amsfonts}       % blackboard math symbols
\usepackage{nicefrac}     
\usepackage{amsmath}% compact symbols for 1/2, etc.
\usepackage{microtype}      % microtypography
\usepackage{lipsum}
\usepackage{fancyhdr}       % header
\usepackage{graphicx}       % graphics
\usepackage{amsmath}
\usepackage{booktabs}
\usepackage{tabularx}
\usepackage{ragged2e}
\usepackage{seqsplit} % enables line breaks inside long "no-space" strings
\usepackage{gensymb}
\usepackage{subcaption}
\usepackage{siunitx}
\usepackage{multirow}
\usepackage{placeins}
\usepackage[normalem]{ulem}
\usepackage{xcolor} % revision markup: \sout{} deleted, \textcolor{blue}{} added
\usepackage{marginnote} % revision markup: notes in the left margin, also inside floats and captions
\ifshowchanges
\NewDocumentCommand{\revmargin}{m}{\leavevmode\marginnote{\footnotesize\color{red}#1}}
\AtBeginDocument{\reversemarginpar\setlength{\marginparwidth}{0.75in}\setlength{\marginparsep}{0.12in}}
\else
\DeclareRobustCommand{\sout}[1]{\ifhmode\ifdim\lastskip>0pt \ignorespaces\fi\fi}
\colorlet{blue}{black}

\NewDocumentCommand{\revmargin}{m}{}
\fi
\ExplSyntaxOn
\seq_new:N \l__revnote_reviewers_seq
\prop_new:N \l__revnote_comments_prop
\tl_new:N \l__revnote_tl
\tl_new:N \l__revnote_out_tl
\cs_new_protected:Npn \__revnote_add:w #1 . #2 \q_stop
  {
    \prop_get:NnNTF \l__revnote_comments_prop {#1} \l__revnote_tl
      { \tl_put_right:Nn \l__revnote_tl { ,~#2 } }
      {
        \seq_put_right:Nn \l__revnote_reviewers_seq {#1}
        \tl_set:Nn \l__revnote_tl {#2}
      }
    \prop_put:NnV \l__revnote_comments_prop {#1} \l__revnote_tl
  }
\NewDocumentCommand{\revnote}{om}
  {
    \seq_clear:N \l__revnote_reviewers_seq
    \prop_clear:N \l__revnote_comments_prop
    \tl_clear:N \l__revnote_out_tl
    \clist_map_inline:nn {#2} { \__revnote_add:w ##1 \q_stop }
    \seq_map_inline:Nn \l__revnote_reviewers_seq
      {
        \tl_if_empty:NF \l__revnote_out_tl { \tl_put_right:Nn \l__revnote_out_tl { \\ } }
        \prop_get:NnN \l__revnote_comments_prop {##1} \l__revnote_tl
        \tl_put_right:Nn \l__revnote_out_tl { Rev.\nobreakspace ##1 \\ Com.\nobreakspace }
        \tl_put_right:NV \l__revnote_out_tl \l__revnote_tl
      }
    \IfValueT {#1} { \tl_put_right:Nn \l__revnote_out_tl { \\ #1 } }
    \exp_args:NV \revmargin \l__revnote_out_tl
  }
\ExplSyntaxOff
\graphicspath{{figures/}}     % organize your images and other figures under media/ folder
\DeclareMathOperator{\atantwo}{atan2}
\DeclareSIUnit{\px}{px}

\title{Meteosat Third Generation imagery improves CNN-based SSI retrieval
}

\author{
  Gordei Pribõtkin \\
  Institute of Computer Science \\
  University of Tartu \\
  STACC OÜ\\
  Tartu, Estonia\\
  \texttt{gordei.pribotkin@ut.ee} \\
   \And
  Piia Post, Velle Toll \\
  Centre for Climate Research, Institute of Physics \\
  University of Tartu \\
  Tartu, Estonia\\
  \texttt{piia.post@ut.ee} \\
  \texttt{velle.toll@ut.ee} \\
}

\begin{document}
\maketitle

\begin{abstract}
Accurate Surface Solar Irradiance (SSI) estimation is increasingly important for photovoltaic energy monitoring and forecasting. The recently introduced Meteosat Third Generation (MTG) satellite constellation provides imaging data with higher spatial resolution compared to the Meteosat Second Generation (MSG) satellite constellation, but its benefits for machine-learning-based SSI retrieval are not well established. We \revnote{1.4, 2.1, 2.2}\sout{introduce a multi-imager and multi-resolution convolutional neural network architecture}\textcolor{blue}{compare multi-resolution convolutional neural network models} for 10-minute SSI retrieval over Northern Europe (Estonia) using \textcolor{blue}{either} MSG/SEVIRI \sout{and}\textcolor{blue}{or} MTG/FCI satellite imagery. We evaluate model performance against ground-based pyranometer measurements from eight Estonian meteorological stations using site-based cross-validation \textcolor{blue}{with a temporally independent test year} and multiple training seeds. Model performance is also compared with the SARAH-3 physics-based satellite SSI product. The \sout{hybrid SEVIRI–FCI}\textcolor{blue}{FCI-only} model significantly outperformed the SEVIRI-only model under overcast and cloudy conditions, reducing RMSE by \sout{\mbox{\SI{8.2}{\watt\per\square\meter}} and \mbox{\SI{5.7}{\watt\per\square\meter}}}\textcolor{blue}{\SI{5.1}{\watt\per\square\meter} and \SI{6.5}{\watt\per\square\meter}}, respectively. However, under partly cloudy or clear skies, there was no statistically significant difference in RMSE between the \sout{SEVIRI-FCI hybrid}\textcolor{blue}{FCI-only} and the SEVIRI-only models. \textcolor{blue}{In all-sky conditions, the RMSE of the FCI-only model was \SI{5.1}{\watt\per\square\meter} lower on average, but this difference was not statistically significant after Bonferroni correction.} Compared with physics-based SARAH-3, the \sout{hybrid}\textcolor{blue}{FCI-only} model yielded skill scores of \sout{\mbox{\SI{35}{\percent}}}\textcolor{blue}{\SI{28}{\percent}} under overcast conditions, \sout{\mbox{\SI{21}{\percent}}}\textcolor{blue}{\SI{19}{\percent}} under cloudy conditions, and \sout{\mbox{\SI{20}{\percent}}}\textcolor{blue}{\SI{17}{\percent}} overall. Furthermore, both models underperformed SARAH-3 in clear-sky conditions. These results show that higher-resolution MTG/FCI imagery improves CNN-based SSI retrieval when clouds dominate irradiance variability, but also indicate that higher resolution alone is insufficient to address clear-sky limitations in machine-learning-based SSI retrieval.
\end{abstract}

% keywords can be removed
%\keywords{Remote Sensing\and Computer Vision}

%\tableofcontents
\newpage
\section{Introduction}
Due to the growing adoption of photovoltaic (PV) energy generation, accurately estimating the amount of solar radiation reaching the surface is increasingly important \cite{solartrends}. Gridded Surface Solar Irradiance (SSI) products have become commonplace for estimating energy production, which is important for PV energy infrastructure planning and monitoring \cite{solar-planning-handbook, PV-faultdetection-ANN}. SSI products are often based on physical models, which rely on explicit physical principles to derive solar irradiance variables from satellite images taken in various visible and infra-red bands. One such method is Heliosat \cite{heliosat}, which is used for the creation of surface solar radiation datasets such as SARAH-3 \cite{sarah3} and the Copernicus Atmosphere Monitoring Service (CAMS) gridded solar radiation service \cite{cams_dataset, cams_guide}, which are the main SSI estimation sources for Europe. SARAH-3 uses data from Meteosat First Generation (MFG) and Meteosat Second Generation (MSG) satellites; CAMS uses data from MSG and Himawari 8 satellites. Similar meteorological products using physical models include the National Solar Radiation Database (NSRDB) \cite{Sengupta2018_NSRDB}, primarily used for North America, and INSAT-3D for the Indian Peninsula \cite{INSAT-3D}. 

Current gridded surface irradiance products have limitations both in horizontal and temporal resolution and accuracy \cite{sarah3, cams_guide}. For SARAH, the spatial resolution is $0.05\degree \times 0.05\degree$, corresponding to about $\SI{5.6}{\km} \times \SI{5.6}{\km}$, while CAMS is coarser, at $0.1\degree \times 0.1\degree$ or approximately $\SI{11}{\km}\times \SI{11}{\km}$. The temporal resolutions of SARAH and CAMS are 30 and 15 minutes, respectively. While CAMS offers SSI time-series predictions for point-locations at a temporal resolution of up to 1 minute, these time-series are interpolated from the gridded product in both time and space, and the underlying model and data are not more accurate than the gridded product.  INSAT-3D and NSRDB offer comparable temporal and spatial resolutions. 

Meteosat Third Generation (MTG) satellite data from the Flexible Combined Imager (FCI) sensor has recently become available, offering a spatial resolution of \SI{0.5}{\km} to \SI{1}{\km} at nadir, depending on the wavelength \cite{EUMETSAT_MTG_FCI_L1C_Data_Guide_v1_8_2026}. For MSG, using the Spinning Enhanced Visible and Infrared Imager (SEVIRI) sensor, spatial resolutions of \SI{1}{\km} nadir for a single visible band channel and \SI{3}{\km} nadir for other visible and infrared (IR) channels are available \cite{eumetsat2017_msg_l15_format}. Both MTG and MSG offer a temporal resolution of at least 10 minutes, with some products offering even higher temporal resolutions. Additionally, the newer FCI imager offers higher spectral resolution compared to SEVIRI. 

Modern machine learning methods, combined with greater spatial and spectral resolutions available with newer EUMETSAT satellites, could yield improvements in accuracy for remote SSI retrieval compared to physical models \cite{fracne-ssi-generalizaiton, central-europe-cnn, eumetsat2024fci}. Machine Learning (ML) methods offer great representation ability and have been used to model physical properties in remote sensing \cite{earth-system-dl}. Neural networks (NNs) have been applied to remote sensing image data for SSI retrieval, showing promising results. For example, for the region of China, single-channel MTSAT images derived from the visible light band have been used with \SI{1}{\km} nadir spatial resolution and hourly average SSI measurements \cite{SSI-CNN-China}. Verbois et al. \cite{fracne-ssi-generalizaiton} considered a simple NN to study the generalization of SSI retrieval models, using MSG data at \SI{3}{\km} nadir resolution and hourly average pyranometer reading data over France. They found that the model outperformed CAMS, but did not generalize well to stations outside of the training region.  Schuurman and Meyer \cite{central-europe-cnn} trained a convolutional neural network (CNN) model to emulate the SARAH SSI dataset and fine-tuned it on 10-minute average pyranometer readings over Germany, the Netherlands, and Switzerland using MSG data similarly at \SI{3}{\km} nadir resolution. The model was more accurate than the SARAH dataset and showed good generalization in regions outside of the training geographic domain, especially under cloudy conditions. Chen et al. \cite{US-CNN-SSI} targeted SSI and Direct Normal Irradiance (DNI) at a 5-minute temporal resolution using images from the GOES-16 geostationary satellite with a spatial resolution of \SI{2}{\km} nadir in North America. Their approach is unique because they used a multi-branch CNN architecture, where each spectral band image is encoded separately. They used SSI and DNI readings from 7 SURFRAD meteorological stations for the year of 2019 and enforced only a temporal test-train split. Chen et al. reported an improvement in SSI and  DNI estimations overall and in cloudy conditions with their approach, compared to the physics-based NSRDB baseline. 

ML-based SSI retrieval does not always outperform physical models. Poor clear-sky performance and generalization were indicated as issues in several ML-based SSI retrieval studies.  Verbois et al. \cite{fracne-ssi-generalizaiton} and Schuurman and Meyer \cite{central-europe-cnn} proposed that under clear-sky conditions, the model's predictions depend more on the surface albedo, while in cloudy conditions, the atmospheric albedo is more important due to clouds obstructing the view of the surface. This has been suggested as the reason for worse generalization under clear-sky conditions, as surface albedo varies more geographically as opposed to atmospheric albedo. Chen et al. \cite{US-CNN-SSI} proposed another explanation for poor clear-sky performance. Namely, under clear skies, some atmospheric factors like aerosols and water vapour are not well-captured in the used spectral bands, but have the most effect on SSI, which might cause an ML-based model to perform worse.

\begin{figure}[h!]
    \centering
    \includegraphics[width=0.7\linewidth]{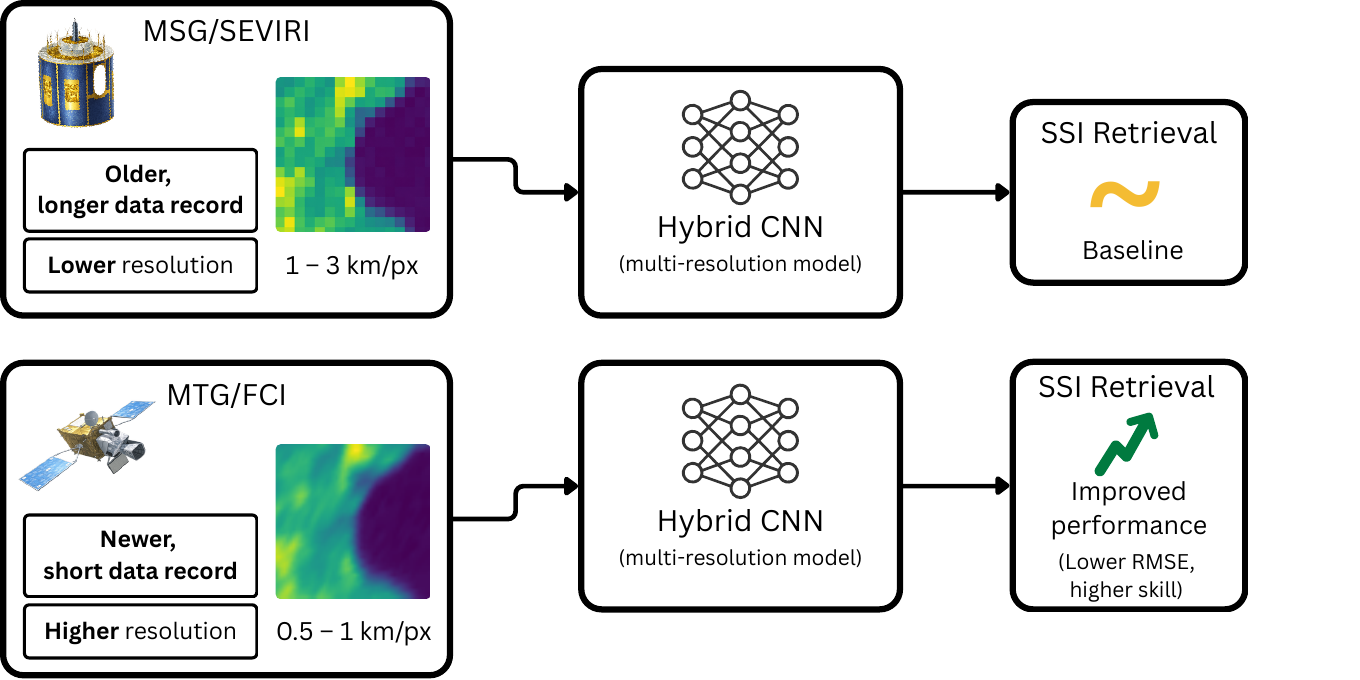}
    \caption{\revnote{2.1}Conceptual study design. We \sout{use the combination of}\textcolor{blue}{compare models using} MSG and MTG data. Our main dataset spans \sout{5 years for MSG/SEVIRI and 1 year for MTG/FCI}\textcolor{blue}{approximately two years for both MSG/SEVIRI and MTG/FCI}. MSG/SEVIRI spatial resolutions are \SI{3}{\km} and \SI{1}{\km} nadir; for MTG/FCI, the spatial resolutions are \SI{1}{\km} and \SI{0.5}{\km} nadir. Images are illustrative. }
    \label{fig:idea_summary}
\end{figure}

In this study, we test whether higher spatial and spectral resolution MTG imagery improves CNN-based SSI retrieval, compared to relying only on MSG data. For this, we \revnote{1.3, 2.1, 2.3}\sout{propose a multi-resolution and multi-imager CNN model}\textcolor{blue}{compare multi-resolution CNN models that use either MSG or MTG imagery} to predict pyranometer readings from meteorological stations in Estonia.  \sout{Our model architecture integrates images from both SEVIRI, at \mbox{\SI{1}{\km}} and \mbox{\SI{3}{\km}} nadir resolution, and, importantly, the higher spatial resolution (\mbox{\SI{0.5}{\km}} and \mbox{\SI{1}{\km}} nadir) FCI images.}\textcolor{blue}{We train the same architecture on SEVIRI images at \SI{1}{\km} and \SI{3}{\km} nadir resolution and, separately, on the higher spatial resolution (\SI{0.5}{\km} and \SI{1}{\km} nadir) FCI images, using the same one-year training period for both models and the following year for testing.} Previous research on ML-based SSI retrieval has used lower-resolution images at \SI{2}{\km}-- \SI{3}{\km} nadir and targeted lower latitudes (Central and Western Europe, China, the United States of America). In addition, related studies mainly provided results for individual models without employing cross-validation or other robust ways to judge the uncertainty of the obtained results \cite{fracne-ssi-generalizaiton, central-europe-cnn, US-CNN-SSI}. We use cross-validation to robustly show that higher-resolution FCI imager data can improve SSI retrieval performance. Furthermore, we find that specifically under clear skies, the predictions of our CNN-based models are more biased, and the training outcomes are much more unstable, when compared to cloudy sky conditions. \autoref{fig:idea_summary} provides a concise graphical summary of our general study design.

\FloatBarrier
\section{Data}

\begin{figure}[h!]
    \centering
    \includegraphics[width=0.6\linewidth]{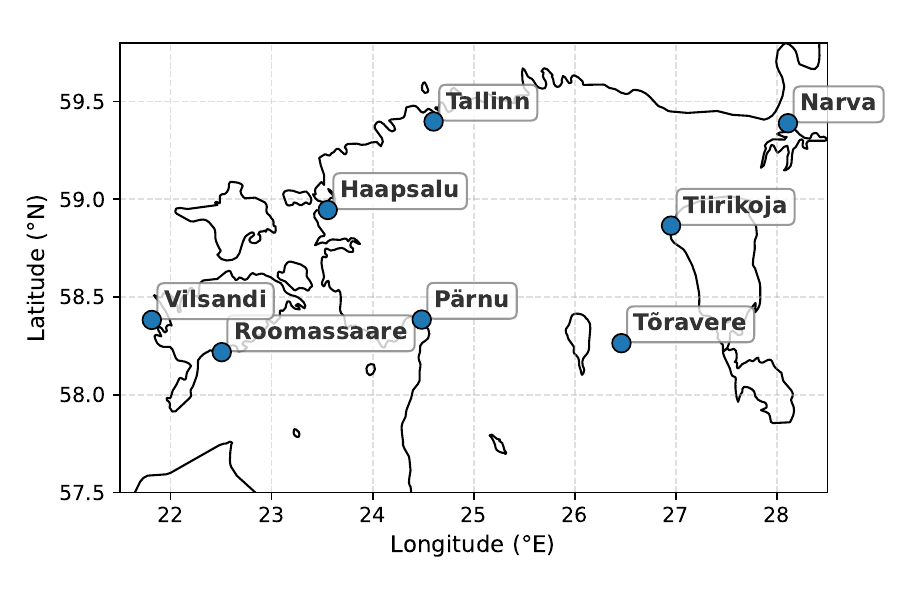}
    \caption{ Map showing locations of meteorological stations.}
    \label{fig:station_locs}
\end{figure}

We use SSI measurements provided by the Estonian Environmental Agency. The measurements originate from 8 meteorological stations in Estonia. All the sites provide quality-controlled 1-minute-average SSI at 10-minute intervals. The station locations are shown in \autoref{fig:station_locs}.\revnote{2.1}\sout{The data that we use spans from 2021.01.01 to 2026.01.01 at each site.}

One of the inputs is images from the SEVIRI imager onboard MSG satellites. We use the Rapid Scan High Rate SEVIRI level 1.5 dataset \cite{EUMETSAT_MSG15_RSS_2009, eumetsat2017_msg_l15_format}. The dataset provides one visible channel with a spatial resolution of \SI{1}{\km} at nadir, and 11 visible and infra-red channels with \SI{3}{\km} nadir resolution. The temporal resolution of the rapid scan rate product is 5 minutes; we sample images at 10-minute intervals to match ground station measurement times. \revnote{1.5, 2.1}\sout{The data we use spans 2021.01.01 - 2026.01.01.} \textcolor{blue}{In March 2026, during our test period, EUMETSAT changed the Level 1.5 calibration of the SEVIRI solar channels, whose calibration coefficients are now updated every repeat cycle instead of every two years \cite{EUMETSAT_SEVIRI_calibration_changes_2026}. We recalibrate the affected images to match the calibration of the training period. This decision was made because our models are trained on images with the earlier calibration, and a change in input calibration between training and testing could bias the SSI estimates of the models that use SEVIRI images. However, as a result, our SEVIRI images from the test period do not follow the updated EUMETSAT calibration.}

We also include FCI imager data from the MTG satellite. We use the FCI Level 1c High Resolution Image Data product, which provides \SI{0.5}{\km} nadir resolution for VIS \SI{0.6}{\micro\meter}, and NIR \SI{2.2}{\micro\meter} channels, and \SI{1}{\km} nadir for IR \SI{3.8}{\micro\meter} and IR \SI{10.5}{\micro\meter} channels \cite{EUMETSAT_EO_EUM_DAT_0665_2024, EUMETSAT_MTG_FCI_L1C_Data_Guide_v1_8_2026}. We sample images with a 10-minute interval, aligning with other datasets. The data is available starting from 2024.09.24. \revnote{2.1}\sout{, but to ensure complete annual coverage and avoid partial-year seasonal bias, we use the period from 2025.01.01 to 2025.12.31.}

The instantaneous SSI values from the SARAH Edition 3 dataset, based on the Heliosat method \cite{heliosat}, are used as a baseline for SSI predictions \cite{sarah3, EUMETSAT_SARAH3_EO_EUM_DAT_0863}. The dataset has a spatial resolution of $0.05\degree \times 0.05\degree$ and a temporal resolution of 30 minutes.  We also include numerical features as inputs to our model. These include latitude and longitude, clear-sky SSI derived from a simple model \cite{Holmgren2018_pvlibpython, ineichen_perez_2002_linke_turbidity}, sine of sun altitude, sine and cosine of sun azimuth, hour of day, and day of year. For the clear-sky SSI we use the Ineichen/Perez model implemented in the \texttt{pvlib} Python library. 

\textcolor{blue}{\revnote{2.1}For our main experiment, we use data from 2024.09.24 to 2026.09.12 from all the aforementioned sources. For the additional experiments with a longer training timespan, we extend our dataset with two additional years of SEVIRI and ground station SSI data, from 2022.09.24 to 2024.09.23.}

We used a geographic site-based split into training, validation, and test sets with cross-validation\revnote{2.1, 2.2}\textcolor{blue}{, combined with a temporal split}. We test on a single site and choose to use two sites for validation to improve model selection for better generalization. \sout{The SEVIRI data in the test sets is additionally limited temporally to the time period of FCI availability so that the data used for testing remains consistent for all models.}\textcolor{blue}{Training and validation sites use data up to 2025.09.23, and we test on data from 2025.09.24 to 2026.09.12.} \sout{Given the limited temporal coverage of the FCI dataset of 12 months, enforcing a temporal train–test split would likely reduce the reliability of our results.} Because solar irradiation exhibits a strong annual cycle, splitting the data into full-year segments \sout{would be preferable, but this is not feasible with the available data at the time our experiments were performed. Separating only by location allows us to have a full year of FCI data for both testing and training.}\textcolor{blue}{is preferable, and the two periods cover approximately one full year each.} However, it must be acknowledged that \sout{leakage between test and train data still exists, stemming from the similarity of general weather patterns due to the shared time period. Consequently, the performance of our model on the test set might be inflated.}\textcolor{blue}{a single year of training data might not represent the interannual variability of weather patterns. Additionally, the validation sites share the training time period, so model checkpoint selection does not account for the temporal shift between training and testing.}

% \begin{figure}[h!]
%     \centering
%     \makebox[\textwidth][c]{%
%     \begin{subfigure}[t]{0.53\textwidth}
%         \centering
%         \includegraphics[width=\linewidth]{figures/stations_lakes.pdf}
%         \caption{}
%         \label{fig:station_locs}
%     \end{subfigure}
%     \hspace{0.02\textwidth}
%     %\hfill
%     \begin{subfigure}[t]{0.53\textwidth}
%         \centering
%         \includegraphics[width=\linewidth]{figures/dataset-spacetime.pdf}
%         \caption{}
%         \label{fig:dataset-spacetime}
%     \end{subfigure}
%     }
%     \caption{Diagrams showing spatial and temporal extent of the dataset. (a) Map showing locations of meteorological stations. (b) Diagram showing temporal data extents for SEVIRI and FCI imagers, and temporal and spatial splits for training, validation, and testing.}
% \end{figure}

% \begin{figure}[h!]
%     \centering
%     \includegraphics[width=0.6\linewidth]{figures/dataset-spacetime.pdf}
%     \caption{Diagram showing temporal data extents for SEVIRI and FCI imagers, and temporal and spatial splits for training, validation, and testing.}
%     \label{fig:station_locs}
% \end{figure}

\section{Model}

%\subsection{Inputs}
    The main inputs of our model are satellite image patches for the regions around each ground station. The physical size of image patches is $\SI{51}{\km}\times\SI{51}{\km}$ for bands with \SI{1}{\km\per\px} and \SI{3}{\km\per\px} resolution and $\SI{25.5}{\km}\times\SI{25.5}{\km}$ for bands with \SI{0.5}{\km\per\px} resolution. To create these patches, we first reproject each original satellite image to a metric grid using EPSG:3301 \cite{epsg3301} coordinate reference system, ensuring that the physical dimensions of each pixel are consistent across the image and between different sites. Then we crop and resample the images to obtain square patches with odd pixel sizes where the site is located exactly in the centre. The physical resolutions for the site patches, in kilometres per pixel, are set to be the same as the nadir resolutions of the source image -- \SI{3}{\km\per\px} to \SI{1}{\km\per\px} for SEVIRI and \SI{1}{\km\per\px} to \SI{0.5}{\km\per\px} for FCI. This means that the images are oversampled due to the oblique viewing angle, but it ensures that as little information as possible is lost. We use the same image size -- \SI{51}{\px} -- for all images, except for SEVIRI \SI{3}{\km} nadir resolution data, where we limit the size to \SI{17}{\px} to \revnote{2.2}\textcolor{blue}{avoid unnecessarily large patch sizes.} The bands and resolutions that we use are summarized in \autoref{tab:imagers-res-bands}. All available FCI spectral bands from the used FCI data product were selected, and we picked a similar set of spectral bands for SEVIRI to make a better comparison between models using different imager combinations. The images are Z-normalized separately for each spectral band by subtracting the training-set mean and dividing by the training-set standard deviation of pixel values. Additionally, our model receives auxiliary numerical inputs:
\begin{itemize}
    \item Latitude and longitude of the site normalized by the geographical bounding box of the region of interest (\SI{57.5}{\degree}\text{N}, \SI{60.9}{\degree}\text{N}, \SI{20.6}{\degree}\text{E}, \SI{29.3}{\degree}\text{E}).
    \item Normalized clear-sky SSI ($\mathrm{SSI}_{cs}$) as $\text{SSI}_{\text{norm}} = \text{SSI}_{cs} / (\SI{800}{\watt\per\square\meter}) $.
    \item Sine and cosine of the phase of day as  $\varphi_{\sin, d} = \sin(2\pi s_d/s_{\text{tot}})$ and $\varphi_{\cos, d} = \cos(2\pi s_d/s_{\text{tot}})$ respectively, where $s_d$ is the number of seconds from midnight at a given timestep and $s_\text{tot} = 86400$ is the total number of seconds in a day.
    \item Sine and cosine of the phase of year as  $\varphi_{ \sin, y} = \sin(2\pi d_y/d_{\text{tot}})$ and $\varphi_{\cos, y} = \cos(2\pi d_y/d_{\text{tot}})$ respectively, where $d_y$ is the number of days from start of the year at a given timestep and $d_\text{tot} = 365$ is the total number of days in a year.
    \item Sine and cosine of the azimuth of the Sun.
    \item Sine of the altitude of the Sun.
\end{itemize}

To align satellite images with ground stations' pyranometer readings, we use the average imaging time band provided with both SEVIRI and FCI data. 

\begin{table*}[h]
\centering
\caption{Imagers, spatial resolutions, and selected spectral bands and their spectral widths.}
\label{tab:imagers-res-bands}
\small
\renewcommand{\arraystretch}{1.1}

\begin{tabularx}{\textwidth}{X X X X p{0.4\textwidth}}
\toprule
Imager & Base resolution & Resampled resolution  & Image size & Band (spectral width) \\
\midrule
SEVIRI & \SI{3}{\km} (nadir) & \SI{3}{\km\per\px}  &\SI{17}{\px} $\times$ \SI{17}{\px} &
VIS \SI{0.6}{\micro\meter} (\SI{0.15}{\micro\meter}), IR \SI{10.8}{\micro\meter} (\SI{2}{\micro\meter}),
IR \SI{3.9} {\micro\meter} (\SI{0.88}{\micro\meter}), IR \SI{1.6} {\micro\meter} (\SI{0.28}{\micro\meter})\\
SEVIRI & \SI{1}{\km} (nadir) & \SI{1}{\km\per\px}   &\SI{51}{\px} $\times$ \SI{51}{\px} & HRV (\SI{0.3}{\micro\meter})\\

FCI & \SI{1}{\km} (nadir) & \SI{1}{\km\per\px}   &\SI{51}{\px} $\times$ \SI{51}{\px} & IR \SI{3.8}{\micro\meter} (\SI{0.4}{\micro\meter}), IR \SI{10.5}{\micro\meter} (\SI{0.7}{\micro\meter}) \\
FCI & \SI{0.5}{\km} (nadir) & \SI{0.5}{\km\per\px} &\SI{51}{\px} $\times$ \SI{51}{\px} & VIS \SI{0.6}{\micro\meter} (\SI{0.05}{\micro\meter}), NIR \SI{2.2}{\micro\meter} (\SI{0.05}{\micro\meter}) \\
\bottomrule
\end{tabularx}
\end{table*}

Additionally, we add trigonometric pixel-coordinate encodings and two-dimensional Gaussian channels to the input site patches. The encodings are based on $x$ and $y$ coordinates of each pixel, which range from -1 to 1, left to right and top to bottom, respectively. The trigonometric coordinate encodings ($E_{\text{sin}}$, $E_{\text{cos}}$) are in the form of the sine and cosine of the angle for each pixel relative to the upward direction on the image; see equations \ref{eq:sine_enc} and \ref{eq:cos_enc}. The Gaussian encoding, $E_{\text{gauss}}$, is calculated according to the Gaussian distribution which is centred in each image; see equation \ref{eq:gauss_enc}.

\begin{align}
    E_{\text{sin}} (x, y) = \sin(\atantwo(x, y)) \label{eq:sine_enc}\\
    E_{\text{cos}} (x, y) = \cos(\atantwo(x, y)) \label{eq:cos_enc}\\
    E_{\text{gauss}} (x, y) = \exp(-(x^2 + y^2)/0.25) \label{eq:gauss_enc}
\end{align}

In our testing, we found that adding coordinate encodings improves the performance of the model and that trigonometric encodings yield better performance compared to raw $x$ and $y$ coordinate encodings without trigonometric transformations.

%\subsection{Target}
\label{sec:target}
The target for the model is a stabilized clear-sky index ($\text{CSI*}$) with an additional term $\varepsilon = \SI{10}{\watt\per\square\meter}$ in the denominator for stability. The target is defined in eq. \ref{eq:target}, where $\text{SSI}$ and $\text{SSI}_{cs}$ are measured and clear-sky irradiances respectively. 
\begin{align}\label{eq:target}
    \text{CSI}^* = \frac{\text{SSI}}{\text{SSI}_{cs} + \varepsilon}
\end{align}

We exclude hours with $\text{SSI}_{cs} < \SI{10}{\watt\per\square\meter}$ from training and evaluation. This means that we train and evaluate only on time steps with sunlight.

\begin{figure}[h]
    \centering
    \includegraphics[width=1\linewidth]{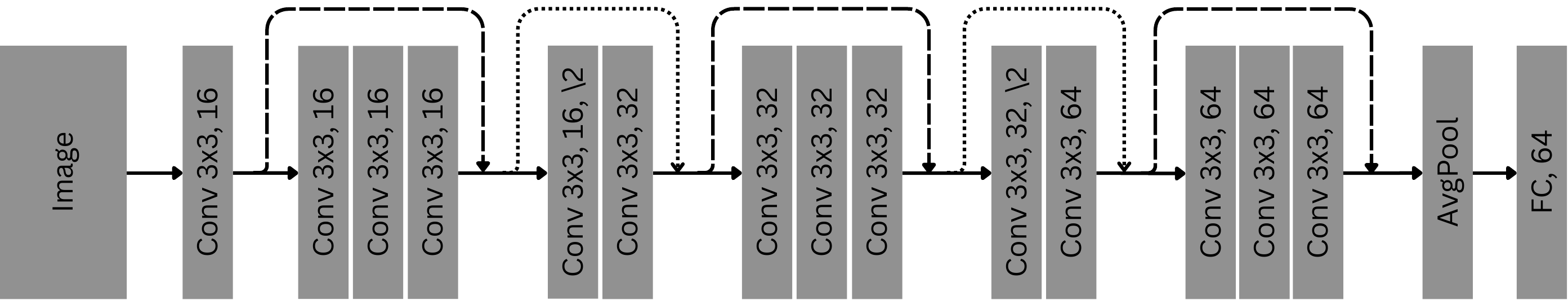}
    \caption{\revmargin{Model size changed with the revision}Diagram of the Branch CNN module. In this diagram, \texttt{Conv 3x3, X} designates a convolutional layer with a $3\times3$ kernel size and X output channels, \texttt{\textbackslash 2} designates downsampling with a stride value of 2 in the convolutional layer. Dashed lines show skip connection without downsampling. Dotted lines show a skip connection that applies downsampling to the residual using a $1\times1$ convolution with a stride value of 2. Before the \texttt{FC} (fully-connected) layer, a two-dimensional adaptive average pool with an output size of 1 is used to reduce the feature map.}
    \label{fig:branch-cnn-arch}
\end{figure}

%\subsection{Architecture}
We use a multi-resolution, multi-branch CNN. The general architecture is similar to architectures proposed in \cite{US-CNN-SSI, branch-multimodal}. In our model, image modalities are separated by the imager and spatial resolution. In total, we have up to 4 modalities -- SEVIRI \SI{3}{\km}, SEVIRI \SI{1}{\km}, FCI \SI{1}{\km} and FCI \SI{0.5}{\km}. Each modality is passed into a separate Branch CNN module, presented in \autoref{fig:branch-cnn-arch}. The Branch CNN follows the ResNet architecture \cite{resnet}, with a similar model being proposed specifically for SSI retrieval in other works \cite{SSI-CNN-China, central-europe-cnn}. We retrieve SSI estimations from the stabilized clear-sky index predicted by the model by solving \autoref{eq:target} for SSI, as described as follows: 
\begin{align}\label{eq:ssi}
    \text{SSI}_{\text{pred}} = \text{CSI}_\mathrm{pred}^* (\text{SSI}_{cs} + \varepsilon).
\end{align} 

\begin{figure}[h]
    \centering
    \includegraphics[width=0.6\linewidth]{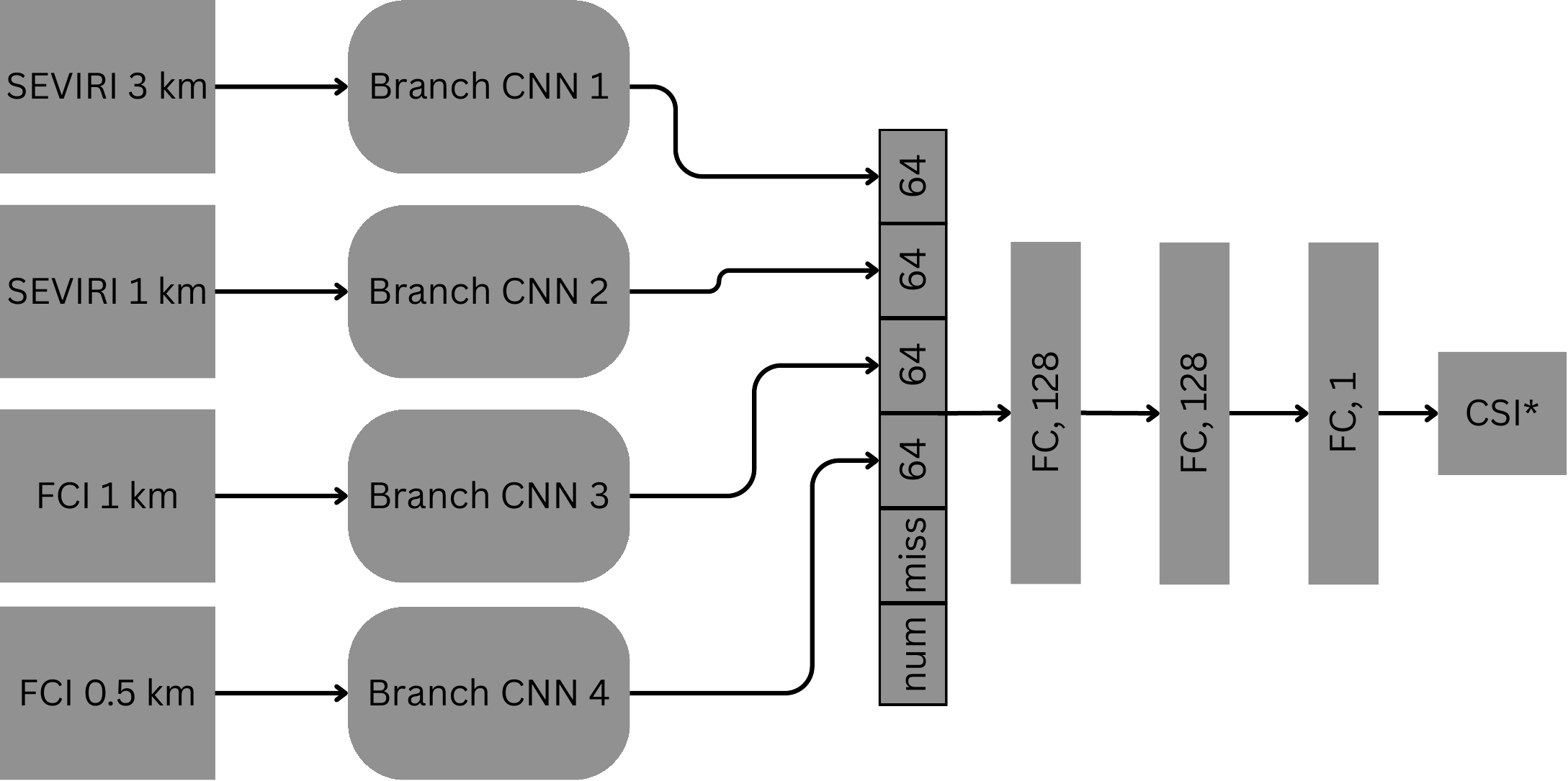}
    \caption{\revmargin{Model size changed with the revision}Diagram of the ML-based SSI retrieval model. \texttt{miss} and \texttt{num} represent missingness flags and numerical features. \texttt{FC, X} designates fully-connected layers with an output size of X.}
    \label{fig:high-level-arch}
\end{figure}

Each branch CNN module produces an embedding of the corresponding image for each modality. The embeddings are then concatenated into a single vector to which the numerical features are appended. The vector is passed to the model head, consisting of fully-connected (FC) layers. The full model architecture is presented in \autoref{fig:high-level-arch}. We include a batch normalization \cite{batchnrom} layer between each convolution and FC layer and every layer uses the Mish \cite{DBLP:journals/corr/abs-1908-08681} activation function. Additionally, the model allows for missing modalities. For that, we append binary missingness flags for each modality (0 - missing, 1 - not missing) to branch embeddings.

\revnote{2.1}\sout{We train two models utilizing the full extent of the SEVIRI data, presented in \mbox{\autoref{tab:model-classes}}. First, we train a model only on SEVIRI with data from the full period in our dataset, from 2021.01.01 to 2025.12.31, resulting in the SEVIRI-only model. To obtain the HYBRID model, we freeze the weights of CNN branches that encode SEVIRI images in the SEVIRI-only model, and train the branches that encode FCI images along with fully connected layers in the head of the model on a full year where FCI is available -- from 2025.01.01 to 2025.12.31. The HYBRID model receives both SEVIRI and FCI images during training, but only FCI branches and head parameters are updated, resulting in the HYBRID model. We chose to freeze the SEVIRI branches during fine-tuning because this allows our model to retain stable representations from the longer SEVIRI dataset when the FCI branches are trained only during the shorter FCI period. We chose to use the FCI data for fine-tuning, as opposed to separately training a two-imager or FCI-only model. This decision was made because, in preliminary testing, we observed the performance outcomes of models trained on a single year of data to be significantly more unstable compared to models using multiple years of SEVIRI data, which would potentially make our results less conclusive. However, this could potentially bias the HYBRID model to perform better in evaluation, as the evaluation data timespan is the same as the fine-tuning data timespan for the HYBRID model. We believe that this effect would be most prominent if the weather patterns in the evaluation year are significantly different compared to the four other years in the SEVIRI-only model training data. Additionally, augmenting the SEVIRI-only model with FCI data is an experiment that is interesting for practical applications, as it allows us to test whether combining abundant data from an older imager with limited data from a newer imager is a viable strategy for improving ML-based SSI retrieval performance.}

We train the models for a maximum of \revnote{2.1}\textcolor{blue}{25} epochs with an early stopping patience of \textcolor{blue}{5} epochs based on the rolling mean of validation RMSE with a window of 5. This metric is also used to select the snapshots of the best model state for both SEVIRI-only and \sout{HYBRID}\textcolor{blue}{FCI-only} models. This allows us to avoid overfitting on the training dataset and potentially obtain the model with the best generalization on unseen data. We also use a ReduceLROnPlateau scheduler with a patience of 3 and a scaling factor of 0.5, which reduces the learning rate during training based on raw validation RMSE. \sout{The SEVIRI-only model is}\textcolor{blue}{Both models are} trained at a learning rate of \num{1e-3}, \sout{the HYBRID model is fine-tuned with a learning rate of \mbox{\num{5e-5}}, and for both models,}\textcolor{blue}{and} we use weight decay of \num{1e-4}, AdamW optimizer, and the Mean Squared Error loss function.  It is also important to mention that the validation RMSE and loss were observed to be somewhat noisy during training.

\textcolor{blue}{\revnote{2.1, 2.3}For our main experiment, we train two models, presented in \autoref{tab:model-classes}. The SEVIRI-only model receives only SEVIRI images, and the FCI-only model receives only FCI images. Both models are trained from randomly initialized weights on the same period, from 2024.09.24 to 2025.09.23, and share the same architecture, numerical inputs, and training procedure. The branches that do not receive input images are inactive and both models have the same number of parameters.  Both models are trained on the same time period to allow us to attribute the difference in performance to the imager rather than the amount of data. }\textcolor{blue}{There is a larger pool of SEVIRI data available and our model architecture supports multi-imager inputs. Following that, we additionally provide results for two-imager hybrid models in \autoref{sec:hybrid-scratch} and models trained on a larger SEVIRI data timespan in \autoref{sec:three-year}}.

\begin{table*}[h!]
\centering
\caption{\revnote{2.1, 2.3}Overview of models used in experiments.}
\label{tab:model-classes}
\begin{tabularx}{0.8\textwidth}{p{2cm} X X}
\toprule
Model class & Training data span  & Number of parameters \\
\midrule
SEVIRI-only& \textcolor{blue}{2024.09.24--2025.09.23} &  482k \\
% &&&\\
\textcolor{blue}{FCI-only}& \textcolor{blue}{2024.09.24--2025.09.23} & 482k \\
\bottomrule
\end{tabularx}
\end{table*}

\section{Evaluation}

\begin{figure}[h!]
    \centering
    \begin{minipage}[t]{0.46\textwidth}
        \vspace{0pt}% top-align both minipages
        \centering
        \captionof{table}{Description of used cross-validation folds. In each fold, all sites other than the test and validation sites are used for training.}
        \label{tab:cv-folds}
        %\small
        \renewcommand{\arraystretch}{1.2}
        \begin{tabularx}{\linewidth}{cc>{\centering\arraybackslash}X}
            \toprule
            Fold & Test site & Validation sites \\
            \midrule
            1 & Haapsalu    & Narva, Pärnu \\
            2 & Tiirikoja   & Haapsalu, Tõravere \\
            3 & Tõravere    & Pärnu, Tallinn \\
            4 & Pärnu       & Roomassaare, Tallinn \\
            5 & Roomassaare & Narva, Vilsandi \\
            6 & Vilsandi    & Roomassaare, Tiirikoja \\
            7 & Tallinn     & Haapsalu, Tõravere \\
            8 & Narva       & Tiirikoja, Vilsandi \\
            \bottomrule
        \end{tabularx}
    \end{minipage}\hfill
    \begin{minipage}[t]{0.50\textwidth}
        \vspace{0pt}
        \normalmarginpar % margin notes of this right-hand minipage go to the right margin, next to the figure
        \centering
        \includegraphics[width=0.9\linewidth]{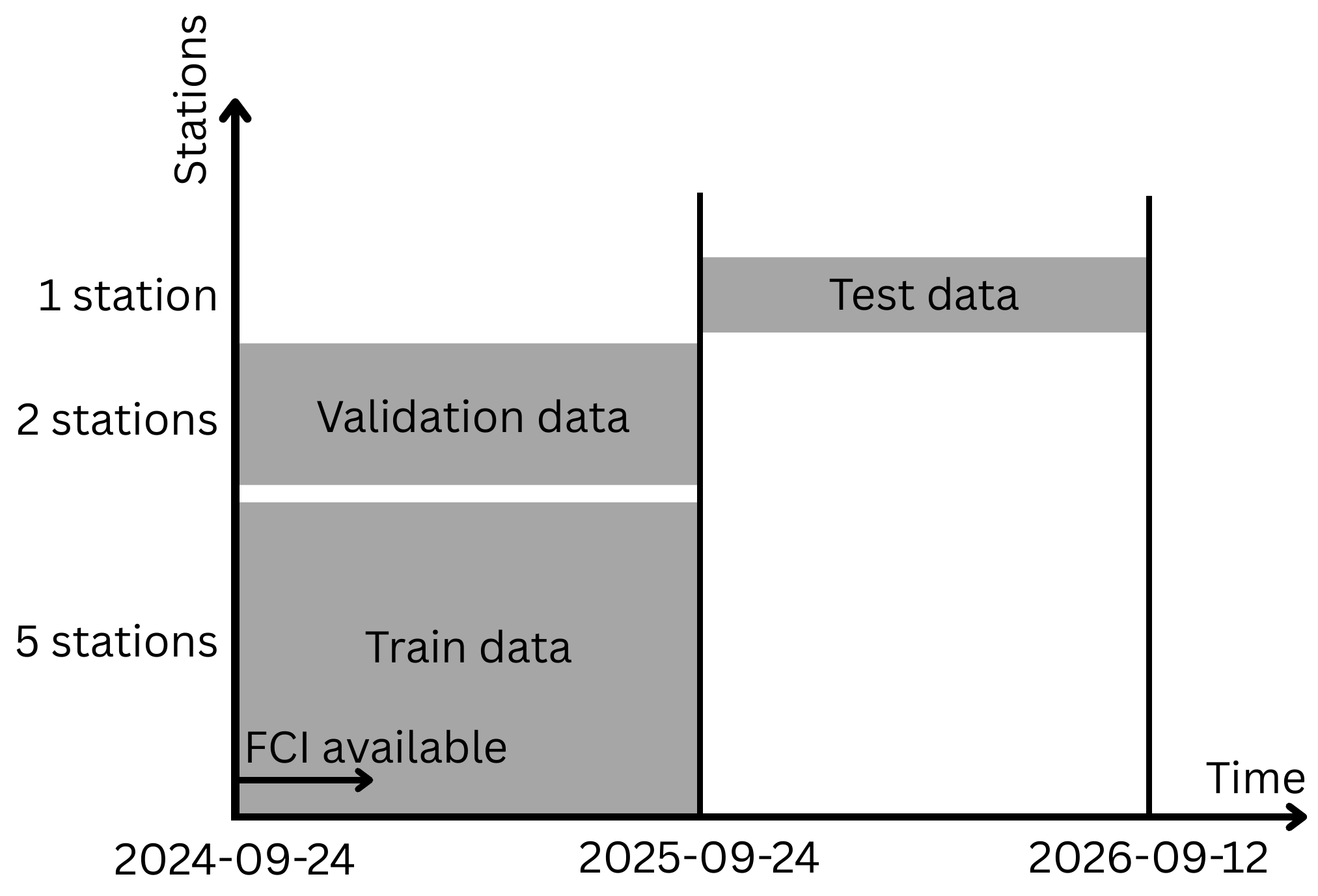}
        \caption{\revnote[Figure updated]{2.1, 2.2}Diagram showing temporal data extents for SEVIRI and FCI imagers, and temporal and spatial splits for training, validation, and testing.}
        \label{fig:dataset-spacetime}
    \end{minipage}
\end{figure}

We use cross-validation to obtain robust estimates of the performance of our models. For that, we permute the train, test and validation sites and train both SEVIRI-only and \revnote{2.1, 2.2}\sout{HYBRID}\textcolor{blue}{FCI-only} models with 5 different seeds for each fold.          \hyperref[tab:cv-folds]{Table~\ref*{tab:cv-folds}} describes the folds used for cross-validation. Testing multiple seeds per fold makes the performance metrics calculated for each fold more accurate and quantifies the variability of training outcomes arising from stochasticity in model initialisation and training. The data split for the folds was chosen in such a way, that each of the eight stations was used once as the held-out test site and twice as a validation site. \textcolor{blue}{The performance of models in each fold is evaluated on the held-out test site on data from a temporally independent period 2025.09.24 -- 2026.09.12. As such, a geographic and temporal split between data observed in training and data used in evaluation is enforced, the test-train split is represented visually in \autoref{fig:dataset-spacetime}. Additionally, in training and testing, we ensure to use the same set of samples for both models.}

% \begin{table*}[h!]
%     \centering
%     \caption{Description of used cross-validation folds.}
%     \begin{tabular}{cccc}
%     \toprule
%          Fold&  Test site&  Validation sites & Train sites\\
%     \midrule
%          1&  Haapsalu&  Narva, Pärnu& Others\\
%          2&  Tiirikoja&  Haapsalu, Tõravere& Others\\
%         3& Tõravere&Pärnu, Tallinn&Others
% \\
     
%  4& Pärnu& Roomassaare, Tallinn&Others\\
%  5& Roomassaare& Narva, Vilsandi&Others
% \\
%  6& Vilsandi& Roomassaare, Tiirikoja&Others\\
%  7& Tallinn& Haapsalu, Tõravere&Others
% \\
%  8& Narva& Tiirikoja, Vilsandi&Others\\
%  \bottomrule
%     \end{tabular}

%     \label{tab:cv-folds}
% \end{table*}
% \begin{figure}[h!]
%     \centering
%     \includegraphics[width=0.6\linewidth]{figures/dataset-spacetime.pdf}
%     \caption{Diagram showing temporal data extents for SEVIRI and FCI imagers, and temporal and spatial splits for training, validation, and testing.}
%     \label{fig:dataset-spacetime}
% \end{figure}

To quantify the performance of our models, we calculate the Mean Absolute Error (MAE), Root Mean Squared Error (RMSE) and Mean Bias Error (MBE), which are defined as follows:
\begin{align}
    \label{eq:MAE}
    \text{MAE} &= \frac{1}{N}\sum_{i=1}^{N}\left|\text{SSI}_{\text{pred},i} - \text{SSI}_i\right|, \\
    \label{eq:RMSE}
    \text{RMSE} &= \sqrt{\frac{1}{N}\sum_{i=1}^{N}\left(\text{SSI}_{\text{pred},i} - \text{SSI}_i\right)^2}, \\
    \label{eq:MBE}
    \text{MBE} &= \frac{1}{N}\sum_{i=1}^{N}\left(\text{SSI}_{\text{pred},i} - \text{SSI}_i\right),
\end{align}
where $N$ is the total number of samples. Additionally, we use skill score (SS) to compare the performance of our models against the SARAH-3 baseline:
\begin{align}
     \label{eq:SS}
    \mathrm{SS} &= 1 - \frac{\text{RMSE}_{\text{model}}}{\text{RMSE}_{\text{baseline}}}.
\end{align}
To obtain sky-condition-specific performance results, the test dataset in each fold is split by sky condition. For that, we use the Variability Index (VI) \cite{variability_index} and daily CSI. The variability index can be used to quantify the variability of the solar radiation on a given day. A higher VI means that SSI values deviated from the clear-sky curve more often and by a higher magnitude, while a lower VI typically means less variability in SSI and/or lower overall SSI magnitude compared to clear-sky SSI. VI value of 1 represents a perfect alignment of observed SSI and clear-sky SSI.  The equations for VI and daily CSI are the following:

\begin{align}
\label{eq:VI}
\mathrm{VI} &= \frac
{
\sum_{i=2}^{n} \sqrt{\left(\text{SSI}_i - \text{SSI}_{i-1}\right)^2 + \Delta t^2}
}
{
\sum_{i=2}^{n} \sqrt{\left(\text{SSI}_{cs, i} - \text{SSI}_{cs, i-1}\right)^2 + \Delta t^2}
},\\ 
\label{eq:daily_CSI}
\mathrm{CSI}_{\mathrm{day}} &=
\frac{\sum_{i=1}^{n} \mathrm{SSI}_{i}}
     {\sum_{i=1}^{n} \mathrm{SSI}_{\mathrm{cs}, i}};
\end{align}
where $\mathrm{SSI}_{i}$ and $\mathrm{SSI}_{\mathrm{cs}, i}$ are the observed SSI and clear-sky model-derived solar irradiances at the $i$-th timestep in a given day, $n$ is the total number of timesteps in that day, and $\Delta t$ is the timestep length in minutes. The definitions for sky conditions\revnote{2.4}\textcolor{blue}{, together with the number of samples in each sky condition in the training and test periods,} are presented in \autoref{tab:sky_condition_definitions}. The definitions for the sky conditions were decided by manually reviewing SSI curves and tuning CSI and VI limits. 

\begin{table}[ht]
\centering
\caption{\revnote{2.4}\textcolor{blue}{Sky condition classification criteria and number of single-timestep samples in test and train periods over all sites. Samples are counted once per site. Test-period samples are counted in the period 2025.09.24--2026.09.12; training-period samples are counted over 2024.09.24--2025.09.23.}}
\label{tab:sky_condition_definitions}
\begin{tabularx}{\textwidth}{p{2cm}p{2.2cm}p{2cm}p{2.2cm}p{2cm}X}
\toprule
Category & Samples in test period& Fractional prevalence & \textcolor{blue}{Samples in training period} & \textcolor{blue}{Fractional prevalence} & Definition \\
\midrule

Clear& \textcolor{blue}{9120}& \textcolor{blue}{5.7\%}& \textcolor{blue}{7575}& \textcolor{blue}{4.7\%}&
$\mathrm{CSI}_{\mathrm{day}} > 0.7$ and $0.8 < VI < 1.2$ \\

Partly cloudy& \textcolor{blue}{38919}&\textcolor{blue}{24.4\%} & \textcolor{blue}{36857}& \textcolor{blue}{22.8\%}&
$\mathrm{CSI}_{\mathrm{day}} > 0.9$ and not a true clear day\\

Cloudy& \textcolor{blue}{66124}& \textcolor{blue}{41.5\%}& \textcolor{blue}{66339}& \textcolor{blue}{41.1\%}&
$0.5 < \mathrm{CSI}_{\mathrm{day}} < 0.9$ and not a true clear day\\

Overcast  & \textcolor{blue}{45070}& \textcolor{blue}{28.3\%}& \textcolor{blue}{50646}& \textcolor{blue}{31.4\%}&
$\mathrm{CSI}_{\mathrm{day}} < 0.5$ and not a true clear day\\
\midrule
 Overall& \textcolor{blue}{159233}& 100\% & \textcolor{blue}{161417}& \textcolor{blue}{100\%} &-\\
 \bottomrule
\end{tabularx}
\end{table}

The final reported performance results are calculated as follows. First, we calculate RMSE, MAE, MBE and SS per each sky condition, seed, fold and model. For each fold, we calculate the average values of these metrics across seeds, producing expected performance per fold for the given sky condition. After that, across the folds, the performance metrics are averaged, the Standard Error of Mean (SEM) is calculated, and statistical tests are performed. 
In addition, we compute seed-wise Relative Standard Deviation of RMSE ($\mathrm{RSD}_{\mathrm{RMSE,seed}}$), which can be expressed as:
\[
\mathrm{RSD_{\mathrm{RMSE,seed}}} = \overline{s_{\mathrm{RMSE}}}/\overline{\mathrm{RMSE}},
\qquad
\overline{s_{\mathrm{RMSE}}} = \frac{1}{F}\sum_{f} s_{\mathrm{RMSE}, f},
% \qquad
% \overline{R} = \frac{1}{F}\sum_{f} \bar{R}_f,
\]
where $\overline{s_{\mathrm{RMSE}}}$ is the mean standard deviation of RMSE across folds, $s_{\mathrm{RMSE},f}$ is the standard deviation of RMSE across seeds for fold $f$, $F$ is the number of folds, and $\overline{\mathrm{RMSE}}$ is the average RMSE across folds and seeds. This metric serves to quantify how much the RMSE of the trained models varies due to the stochasticity of model initialization (different seeds in each fold) on average across folds.

%we calculate the Relative Standard Deviation (RSD) for RMSE across seeds and then average it across folds. This serves as a metric of performance outcome instability stemming from the stochasticity in model initialisation and training. 

\section{Results}
\begin{figure}[h!]
    \centering
    \includegraphics[width=0.5\linewidth]{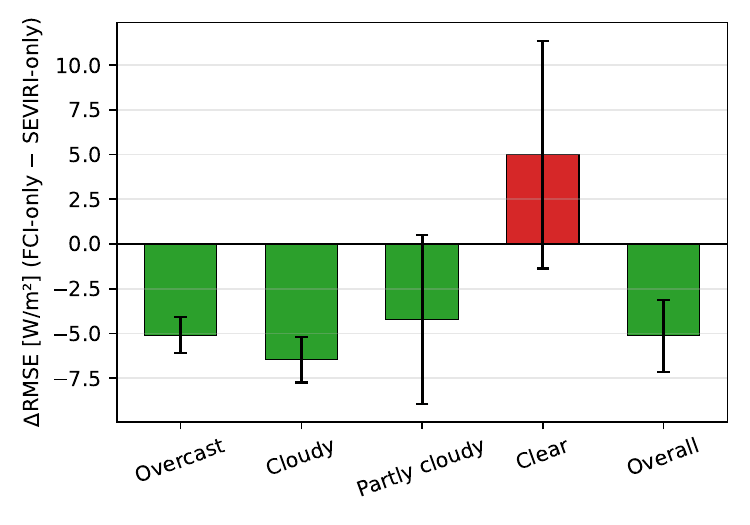}
    \caption{\revnote{2.1}Paired RMSE difference (\sout{HYBRID}\textcolor{blue}{FCI-only} -- SEVIRI-only) by sky condition and overall. Lower value indicates better \sout{HYBRID}\textcolor{blue}{FCI-only} performance when compared to SEVIRI-only (fold-level mean $\pm$ SEM; seeds averaged within each fold).}
    \label{fig:deltaRMSE}
\end{figure}

The \revnote{2.1, 2.2}\sout{HYBRID}\textcolor{blue}{FCI-only} model generally outperforms the SEVIRI-only model. \autoref{tab:cv-hybrid-vs-seviri} shows the fold-level mean RMSE of SEVIRI-only and \sout{HYBRID}\textcolor{blue}{FCI-only} models along with the difference in average RMSE and the p-values of a Bonferroni-corrected two-sided paired t-test on model RMSE values. Additionally,  \autoref{fig:deltaRMSE} shows the average RMSE differences in a graphical form. \textcolor{blue}{Per-fold RMSE and MBE values for each held-out test site, computed on the timestamps where SARAH-3 is available, are listed in \autoref{tab:cv-rmse-perfold-by-sky} and \autoref{tab:cv-mbe-perfold-by-sky}.} Based on the p-values, we can conclude that the \sout{HYBRID}\textcolor{blue}{FCI-only} model outperforms the SEVIRI-only model to a statistically significant degree in overcast\sout{,} \textcolor{blue}{and} cloudy\sout{, and all-sky} conditions with \sout{\mbox{\SI{8.2}{\watt\per\square\metre}}, \mbox{\SI{5.7}{\watt\per\square\metre}}, and \mbox{\SI{4.9}{\watt\per\square\metre}}}\textcolor{blue}{\SI{5.1}{\watt\per\square\meter} and \SI{6.5}{\watt\per\square\meter}} RMSE improvement, respectively. \sout{This shows that our model successfully integrates complementary data from the FCI imager.}\sout{The performance difference in partly cloudy conditions remains small and, in clear-sky conditions, the RMSE of the SEVIRI-only model is, on average, lower than for the two-imager model}\textcolor{blue}{In partly cloudy and all-sky conditions, the RMSE of the FCI-only model is, on average, also lower, by \SI{4.2}{\watt\per\square\meter} and \SI{5.1}{\watt\per\square\meter}, respectively, while in clear-sky conditions, the RMSE of the SEVIRI-only model is, on average, lower, by \SI{5.0}{\watt\per\square\meter}}; however, these differences are not statistically significant. It can be assumed that higher-resolution FCI imager data contains \sout{additional}\textcolor{blue}{more} useful information \textcolor{blue}{for SSI retrieval than SEVIRI imager data} under cloudy skies, but yields no \sout{complementary information}\textcolor{blue}{advantage} for SSI estimation under clearer skies. 

\begin{table}[h!]
\centering
\caption{\revnote{2.1}SSI estimation RMSEs for \textcolor{blue}{FCI-only} and SEVIRI-only models by sky condition. Values are fold-level mean RMSEs $\pm$ SEM. $\Delta$RMSE is the paired fold-level mean of (\textcolor{blue}{FCI-only} $-$ SEVIRI-only) RMSE; negative favours \textcolor{blue}{FCI-only}. \textcolor{blue}{$p$ is the uncorrected $p$-value and} $p_\mathrm{bonf}$ is the Bonferroni-corrected $p$-value (5 tests) from a two-sided paired $t$-test on the same paired differences.}
\label{tab:cv-hybrid-vs-seviri}

\setlength{\tabcolsep}{10pt}
\begin{tabularx}{\textwidth}{X X X l p{1cm} p{1cm}}
\toprule
Condition & SEVIRI-only RMSE [W/m$^2$] & \textcolor{blue}{FCI-only} RMSE [W/m$^2$]
          & $\Delta$ RMSE [W/m$^2$] & \textcolor{blue}{$p$} & $p_\mathrm{bonf}$ \\
\midrule
Overcast & \textcolor{blue}{$59.58 \pm 2.08$} & \textcolor{blue}{$54.48 \pm 2.17$} & \textcolor{blue}{$-5.10 \pm 1.00$} & \textcolor{blue}{0.0014} & \textcolor{blue}{0.0068} \\
Cloudy & \textcolor{blue}{$103.41 \pm 1.56$} & \textcolor{blue}{$96.94 \pm 1.93$} & \textcolor{blue}{$-6.47 \pm 1.27$} & \textcolor{blue}{0.0014} & \textcolor{blue}{0.0072} \\
Partly cloudy & \textcolor{blue}{$83.52 \pm 3.18$} & \textcolor{blue}{$79.30 \pm 3.83$} & \textcolor{blue}{$-4.21 \pm 4.72$} & \textcolor{blue}{0.4021} & 1.0000 \\
Clear & \textcolor{blue}{$44.33 \pm 6.91$} & \textcolor{blue}{$49.34 \pm 4.99$} & \textcolor{blue}{$+5.00 \pm 6.37$} & \textcolor{blue}{0.4580} & 1.0000 \\
\midrule
Overall & \textcolor{blue}{$85.52 \pm 1.78$} & \textcolor{blue}{$80.38 \pm 2.19$} & \textcolor{blue}{$-5.14 \pm 2.00$} & \textcolor{blue}{0.0367} & \textcolor{blue}{0.1833} \\
\bottomrule
\end{tabularx}
\end{table}

Our ML-based models outperform the SARAH-3 baseline under cloudy conditions, see \autoref{tab:cv-vs-sarah}. \revnote{2.1, 2.4}\sout{Both SEVIRI-only and HYBRID models outperform}\textcolor{blue}{The FCI-only model outperforms} SARAH-3 in overcast, cloudy and all-sky conditions\sout{, with the HYBRID model achieving} \textcolor{blue}{with} skill scores of \sout{\mbox{\SI{35}{\percent}}, \mbox{\SI{21}{\percent}} and \mbox{\SI{20}{\percent}}}\textcolor{blue}{\SI{28}{\percent}, \SI{19}{\percent} and \SI{17}{\percent}}, respectively\textcolor{blue}{, while the SEVIRI-only model outperforms SARAH-3 to a statistically significant degree only in overcast and cloudy conditions, with skill scores of \SI{21}{\percent} and \SI{14}{\percent}, respectively}. The difference in performance between our models and the baseline on partly cloudy and clear days is not statistically significant. \textcolor{blue}{However, under clear skies, both models have higher RMSE than SARAH-3 at each of the eight test sites (\autoref{tab:cv-rmse-perfold-by-sky}).} In addition, a negative skill score in clear conditions of approximately \sout{\mbox{\SI{-38}{\percent}} for both models}\textcolor{blue}{\SI{-35}{\percent} and \SI{-54}{\percent} for the SEVIRI-only and FCI-only models, respectively,} is in alignment with previous research retrieving SSI from geostationary satellite data with neural-network-based models \textcolor{blue}{\cite{fracne-ssi-generalizaiton, central-europe-cnn, US-CNN-SSI}}. As physically expected, the SSI prediction for cloudy and partly cloudy conditions is most challenging for all prediction models evident by higher overall RMSE. 

\begin{table}[h!]
\centering
\caption{\revnote{2.1}Per-model performance against SARAH-3 by sky condition. Only timestamps where SARAH-3 is available are included. SARAH-3 RMSEs and skill
scores are fold-level means $\pm$ SEM. $p_\mathrm{bonf}$ is the Bonferroni-corrected (10 tests) $p$-value from a two-sided paired $t$-test on fold-level (model RMSE $-$ SARAH-3 RMSE).}
\label{tab:cv-vs-sarah}
\setlength{\tabcolsep}{10pt}
\begin{tabularx}{\textwidth}{X c c c c c}
\toprule
& \multicolumn{1}{c}{SARAH-3} 
& \multicolumn{2}{c}{SEVIRI-only} 
& \multicolumn{2}{c}{\textcolor{blue}{FCI-only}} \\
\cmidrule(lr){2-2} \cmidrule(lr){3-4} \cmidrule(lr){5-6}
Condition & RMSE [W/m$^2$] 
          & Skill [\%] & $p_\mathrm{bonf}$
          & Skill [\%] & $p_\mathrm{bonf}$ \\
\midrule
Overcast & \textcolor{blue}{$75.24 \pm 1.60$} & \textcolor{blue}{$21.30 \pm 2.15$} & 0.0002 & \textcolor{blue}{$28.00 \pm 2.07$} & $<\!10^{-4}$ \\
Cloudy & \textcolor{blue}{$121.25 \pm 3.10$} & \textcolor{blue}{$13.88 \pm 2.07$} & \textcolor{blue}{0.0030} & \textcolor{blue}{$19.46 \pm 1.78$} & \textcolor{blue}{0.0003} \\
Partly cloudy & \textcolor{blue}{$82.69 \pm 3.99$} & \textcolor{blue}{$-4.62 \pm 10.05$} & 1.0000 & \textcolor{blue}{$2.70 \pm 4.20$} & 1.0000 \\
Clear & \textcolor{blue}{$31.93 \pm 1.33$} & \textcolor{blue}{$-35.43 \pm 14.03$} & \textcolor{blue}{0.6269} & \textcolor{blue}{$-53.77 \pm 14.26$} & \textcolor{blue}{0.0697} \\
\midrule
Overall & \textcolor{blue}{$97.29 \pm 2.28$} & \textcolor{blue}{$11.25 \pm 3.36$} & \textcolor{blue}{0.0858} & \textcolor{blue}{$16.90 \pm 2.17$} & \textcolor{blue}{0.0015} \\
\bottomrule
\end{tabularx}
\end{table}

\begin{figure}[h!]
    \centering

    \begin{subfigure}[t]{0.98\textwidth}
        \centering
        \includegraphics[width=\linewidth]{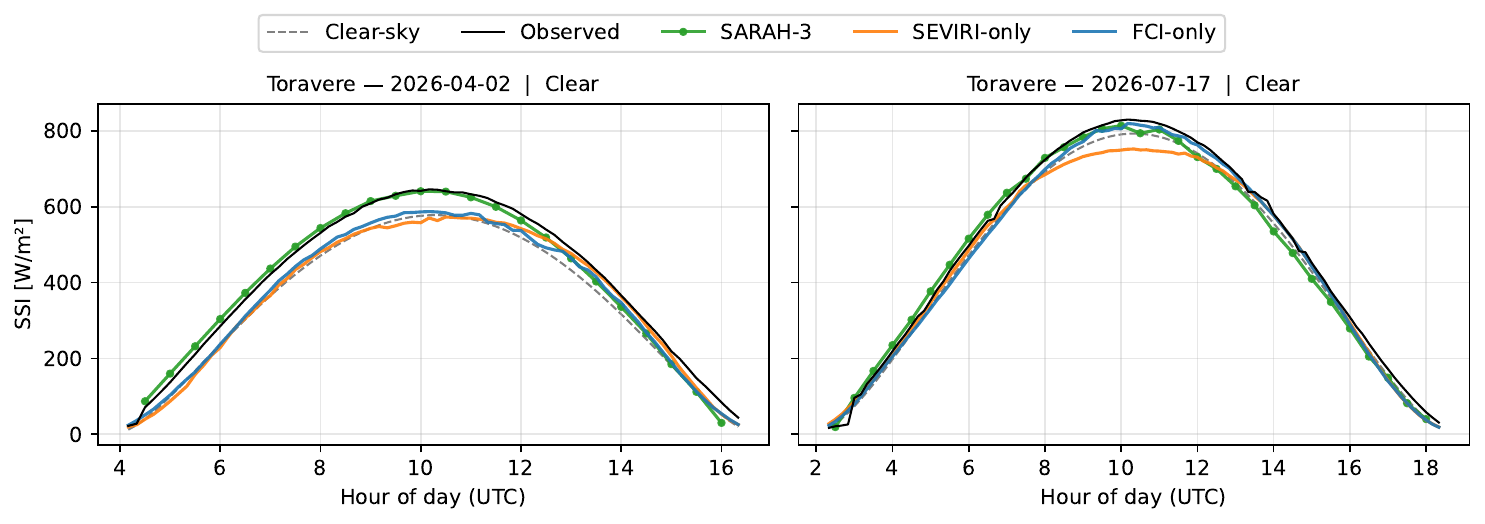}
        \caption{}
        \label{fig:tallinn_cs_samples}
    \end{subfigure}

    \vspace{0.5em}

    \begin{subfigure}[t]{0.98\textwidth}
        \centering
        \includegraphics[width=\linewidth]{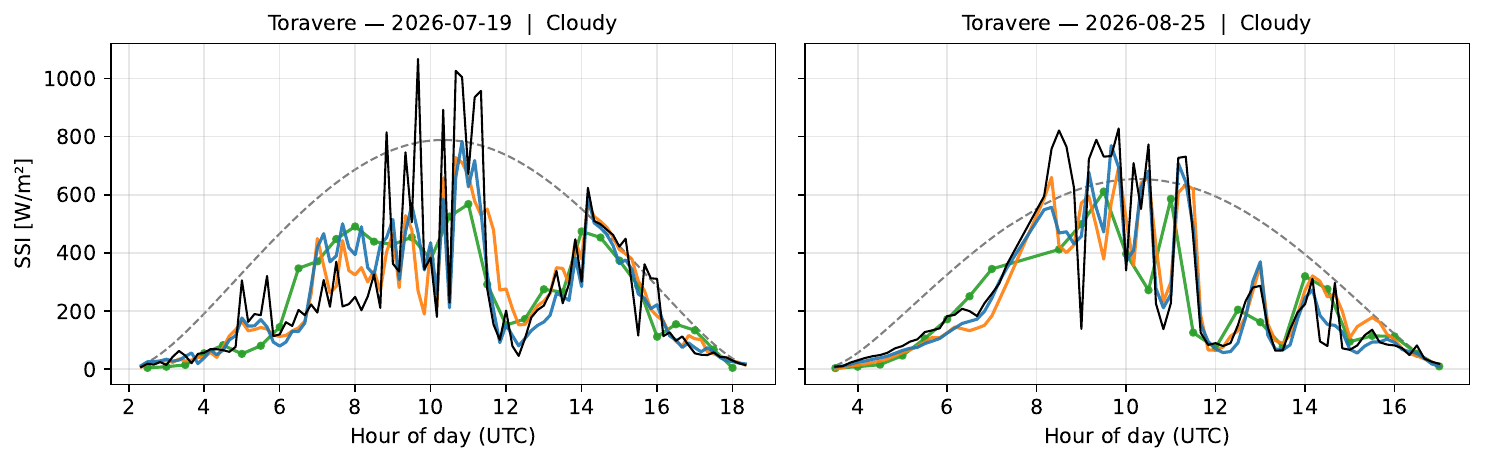}
        \caption{}
        \label{fig:tallinn_cloudy_samples}
    \end{subfigure}

    \caption{\revnote[Figure updated]{2.1}Example clear-sky and cloudy-sky samples for Tõravere. Models were taken from one of the seeds and the fold where the Tõravere site was the test site. (a) Clear days, (b) Cloudy days.}
    \label{fig:tallinn_samples}
\end{figure}

We find that bias accounts for a large portion of prediction errors for both \revnote{2.1}\sout{HYBRID}\textcolor{blue}{FCI-only} and SEVIRI-only models in clear and partly cloudy conditions. \autoref{tab:cv-bias-share} presents the bias share of error for the SARAH-3 baseline and our models. The bias share of error indicates what proportion of MAE can be explained by the bias of the model. In clear conditions, for both models, the bias share is higher, approximately \sout{0.6}\textcolor{blue}{0.6 for the SEVIRI-only model and 0.7 for the FCI-only model}, when compared to the overall bias share of \sout{0.23}\textcolor{blue}{0.22 and 0.24, respectively}. Additionally, the overall bias share of our models -- \sout{0.23}\textcolor{blue}{0.22 and 0.24} is much higher compared to the overall bias share of SARAH-3 -- \sout{0.06}\textcolor{blue}{0.08}. We conclude that the bias of our models is likely a large factor contributing to poorer comparative performance in clear-sky conditions. As an illustrative example, \autoref{fig:tallinn_samples} presents prediction samples for the Tõravere site for clear and cloudy days. For one of the clear days (2026-04-02), a large bias in SSI predictions of both \sout{SEVIRI and HYBRID}\textcolor{blue}{SEVIRI-only and FCI-only} models can be clearly observed. 

Clear-sky performance is not only poorer but also more variable for our ML-based prediction, compared to other sky conditions, likely due to stochasticity in model initialization. To quantify this, \autoref{tab:cv-bias-share} presents the \revnote{2.1}\sout{fold-level average of the seed-level}\textcolor{blue}{seed-wise} relative standard deviation (RSD) of RMSE for our models\textcolor{blue}{, which is the across-seed standard deviation of RMSE averaged over folds and divided by the mean RMSE}. In clear-sky conditions, the seed-wise RSD of RMSE is \sout{close to \mbox{\SI{30}{\percent}} for both HYBRID and SEVIRI models}\textcolor{blue}{\SI{32}{\percent} for the FCI-only model and \SI{30}{\percent} for the SEVIRI-only model}, while for all conditions together, the training outcomes are more stable with RSD of RMSE of just \sout{\mbox{\SI{4}{\percent}} -- \mbox{\SI{5}{\percent}}}\textcolor{blue}{\SI{6}{\percent} for both models}. It is important to note that this standard deviation represents the variation in performance on the same fold, so the test site variation does not affect it. 

\begin{table}[h!]
\centering
\caption{\revnote{2.1}Bias share of the error, $|\mathrm{MBE}|/\mathrm{MAE}$ (indicates what proportion of MAE can be explained by the bias of the model), by sky
condition. Only timestamps where SARAH-3 is available are included. Values are fold-level mean $\pm$ SEM, bounded in $[0,1]$
($0$ = zero-mean residuals, $1$ = pure bias).
$\mathrm{RSD}_\mathrm{RMSE, seed}$ (Relative Standard Deviation) is the mean within-fold across-seed std of
RMSE divided by the fold-level mean RMSE. \sout{Both are descriptive diagnostics}\textcolor{blue}{The RSD is a descriptive diagnostic} of optimization noise (not applicable to SARAH-3, which has no training seed).}
\label{tab:cv-bias-share}

\begin{tabularx}{\textwidth}{X c c c c c}
\toprule
& \multicolumn{1}{c}{SARAH-3}
& \multicolumn{2}{c}{SEVIRI-only}
& \multicolumn{2}{c}{\textcolor{blue}{FCI-only}} \\
\cmidrule(lr){2-2} \cmidrule(lr){3-4} \cmidrule(lr){5-6}
Condition
& $|\mathrm{MBE}|/\mathrm{MAE}$
& $|\mathrm{MBE}|/\mathrm{MAE}$ & $\mathrm{RSD}_\mathrm{RMSE, seed}$
& $|\mathrm{MBE}|/\mathrm{MAE}$ & $\mathrm{RSD}_\mathrm{RMSE, seed}$ \\
\midrule
Overcast & \textcolor{blue}{$0.230 \pm 0.022$} & \textcolor{blue}{$0.322 \pm 0.033$} & \textcolor{blue}{$0.109$} & \textcolor{blue}{$0.303 \pm 0.028$} & \textcolor{blue}{$0.095$} \\
Cloudy & \textcolor{blue}{$0.118 \pm 0.018$} & \textcolor{blue}{$0.188 \pm 0.028$} & \textcolor{blue}{$0.055$} & \textcolor{blue}{$0.211 \pm 0.033$} & \textcolor{blue}{$0.049$} \\
Partly cloudy & \textcolor{blue}{$0.182 \pm 0.043$} & \textcolor{blue}{$0.350 \pm 0.060$} & \textcolor{blue}{$0.110$} & \textcolor{blue}{$0.364 \pm 0.036$} & \textcolor{blue}{$0.103$} \\
Clear & \textcolor{blue}{$0.309 \pm 0.048$} & \textcolor{blue}{$0.600 \pm 0.069$} & \textcolor{blue}{$0.304$} & \textcolor{blue}{$0.700 \pm 0.069$} & \textcolor{blue}{$0.317$} \\
\midrule
Overall & \textcolor{blue}{$0.079 \pm 0.025$} & \textcolor{blue}{$0.215 \pm 0.033$} & \textcolor{blue}{$0.062$} & \textcolor{blue}{$0.243 \pm 0.025$} & \textcolor{blue}{$0.063$} \\
\bottomrule
\end{tabularx}
\end{table}

\textcolor{blue}{\revnote{2.1, 2.3}To test whether combining the two imagers or a longer SEVIRI training record changes these results, we additionally trained a HYBRID model, which receives both SEVIRI and FCI images and is trained from scratch on the same year as the SEVIRI-only and FCI-only models (\autoref{sec:hybrid-scratch}), and three models with SEVIRI pre-training on three years, from 2022.09.24 to 2025.09.23 (\autoref{sec:three-year}). Of the latter, the SEVIRI-only-3Y model is not fine-tuned, while the SEVIRI-FN-3Y and HYBRID-FN-3Y models are fine-tuned on the year preceding the test period, without and with added FCI branches, respectively.}

\textcolor{blue}{\revnote{2.1, 2.3}None of the additional models has a lower all-sky RMSE than the FCI-only model, although the SEVIRI-FN-3Y model comes close. \autoref{tab:cv-rmse-3y} lists the RMSE of all models by sky condition and overall, and \autoref{tab:cv-overall-hybrid-scratch} lists further all-sky metrics of the HYBRID model. The HYBRID model has an all-sky RMSE of \SI{81.38}{\watt\per\square\meter}, which is \SI{4.1}{\watt\per\square\meter} lower than that of the SEVIRI-only model, but \SI{1.0}{\watt\per\square\meter} higher than that of the FCI-only model. The SEVIRI-only-3Y model has \SI{1.0}{\watt\per\square\meter} lower all-sky RMSE than the SEVIRI-only model, and fine-tuning lowers it by a further \SI{3.6}{\watt\per\square\meter}, to \SI{81.02}{\watt\per\square\meter} for the SEVIRI-FN-3Y model, which is \SI{0.6}{\watt\per\square\meter} higher than the all-sky RMSE of the FCI-only model. The HYBRID-FN-3Y model has an all-sky RMSE of \SI{81.61}{\watt\per\square\meter}, which is \SI{0.6}{\watt\per\square\meter} higher than that of the SEVIRI-FN-3Y model. We did not test these differences for statistical significance, so they are reported as descriptive only. It can be assumed that adding a second imager does not lower the all-sky RMSE in our setting, and that most of the improvement of the SEVIRI-FN-3Y model comes from both the fine-tuning stage and the longer initial training period.}

% \begin{table}[h!]
% \centering
% \caption{Paired $\Delta$RMSE (HYBRID $-$ SEVIRI) by sky condition and overall.
% Negative values favour HYBRID. $p_\mathrm{bonf}$ is the Bonferroni-corrected
% $p$-value (across conditions) from a two-sided paired $t$-test on fold-level differences.}
% \label{tab:cv-delta-rmse}
% \begin{tabular}{l c c c c}
% \toprule
% Condition & $n$ & $\Delta$RMSE [W/m$^2$] & 95\% CI & $p_\mathrm{bonf}$ \\
% \midrule
% Overcast & 8 & $-8.17 \pm 1.32$ & $[-11.29,\,-5.05]$ & 0.0022 \\
% Cloudy & 8 & $-5.67 \pm 0.75$ & $[-7.43,\,-3.90]$ & 0.0006 \\
% Partly cloudy & 8 & $-1.34 \pm 1.36$ & $[-4.55,\,+1.87]$ & 1.0000 \\
% Clear & 8 & $+0.70 \pm 2.27$ & $[-4.67,\,+6.07]$ & 1.0000 \\
% Overall & 8 & $-4.94 \pm 0.77$ & $[-6.76,\,-3.12]$ & 0.0018 \\
% \bottomrule
% \end{tabular}
% \end{table}
  
\FloatBarrier
\section{Discussion}

We found that the model using \revnote{1.7, 2.1}\sout{both FCI and SEVIRI imagers}\textcolor{blue}{only FCI images} consistently outperforms a model using only SEVIRI images in overcast\sout{,} \textcolor{blue}{and} cloudy\sout{ and all-sky} conditions by approximately \sout{\mbox{\SI{15}{\percent}}, \mbox{\SI{6}{\percent}} and \mbox{\SI{6}{\percent}}}\textcolor{blue}{\SI{9}{\percent} and \SI{6}{\percent}}, respectively. However, no statistically significant difference was observed in partly cloudy\sout{ and}\textcolor{blue}{,} clear \textcolor{blue}{and all-sky} conditions\textcolor{blue}{, although the all-sky RMSE of the FCI-only model is, on average, approximately \SI{6}{\percent} lower}. This suggests that our model was able to utilize higher-resolution FCI imager data to better resolve the effects of finer cloud features on SSI. However, when clouds have a smaller effect on SSI, the FCI data did not \sout{provide any additional useful information to the model}\textcolor{blue}{give the model a measurable advantage}. \sout{the approach is successful in fusing data from two imagers, combining the longer timespan of SEVIRI data and the higher spatial and spectral resolution of FCI data.}\textcolor{blue}{In general, FCI imagery provides a better basis for CNN-based SSI retrieval than SEVIRI imagery under overcast and cloudy skies.}

\textcolor{blue}{\revnote{2.1, 2.3}The additional models show that combining the two imagers did not improve all-sky performance, which suggests that SEVIRI images add little information to FCI images for SSI retrieval in our setting. Furthermore, a SEVIRI-only model pre-trained on three years and fine-tuned on the year preceding the test period closes most of the all-sky RMSE gap to the FCI-only model. The FCI-only model retains the lowest all-sky RMSE of all our models, approximately \SI{6}{\percent} lower than that of the SEVIRI-only model, but less than \SI{1}{\percent} lower than that of the SEVIRI-FN-3Y model. As such, the advantage of FCI imagery is clearest when both imagers are trained on the same period, and additional training years also likely have a positive effect on model performance. It is unclear why fine-tuning on the year preceding the test period improves performance in the case of SEVIRI-FN-3Y. A possible reason is that the fine-tuning stage gives more weight to the year closest to the test period and that year shares similar weather patterns to the year used for testing. However, we did not test this explanation, and the differences between these models were not tested for statistical significance.}

Our models generally outperform SARAH-3 in more cloudy conditions. The \revnote{2.1, 2.4}\sout{HYBRID}\textcolor{blue}{FCI-only} model yielded a statistically significant improvement over the baseline overall with a \sout{\mbox{\SI{20}{\percent}}}\textcolor{blue}{\SI{17}{\percent}} skill score and skill scores of \sout{\mbox{\SI{35}{\percent}}}\textcolor{blue}{\SI{28}{\percent}}  and \sout{\mbox{\SI{21}{\percent}}}\textcolor{blue}{\SI{19}{\percent}} in overcast and cloudy conditions\textcolor{blue}{, while the SEVIRI-only model yielded a statistically significant improvement only in overcast and cloudy conditions, with skill scores of \SI{21}{\percent} and \SI{14}{\percent}, respectively}. However, in other sky conditions, we did not observe a statistically significant difference in performance. \sout{Furthermore,}\textcolor{blue}{Nevertheless,} in clear conditions, our models were \textcolor{blue}{worse compared to the physics-based SARAH-3 baseline at each of the eight test sites, with RMSE}, on average, \sout{\mbox{\SI{38}{\percent}} worse compared to the physics-based SARAH-3 baseline}\textcolor{blue}{\SI{35}{\percent} and \SI{54}{\percent} higher for the SEVIRI-only and FCI-only models, respectively}. There is evidence to suggest that the main factor in poorer clear-sky performance is model bias, as the respective bias share of error is higher -- \sout{\mbox{\SI{60}{\percent}}}\textcolor{blue}{\SI{60}{\percent} for the SEVIRI-only model and \SI{70}{\percent} for the FCI-only model}, when compared to values of \sout{\mbox{\SI{20}{\percent}} -- \mbox{\SI{30}{\percent}}}\textcolor{blue}{\SI{19}{\percent} -- \SI{32}{\percent} and \SI{21}{\percent} -- \SI{30}{\percent}, respectively,} in cloudy and overcast conditions. Another issue with our models in clear-sky conditions was the comparatively high instability of performance outcomes across different seeds. The across-seed relative standard deviation of RMSE was \sout{approximately \mbox{\SI{30}{\percent}}}\textcolor{blue}{\SI{32}{\percent} for the FCI-only model and \SI{30}{\percent} for the SEVIRI-only model}, compared to values of \sout{\mbox{\SI{3}{\percent}} - \mbox{\SI{10}{\percent}}}\textcolor{blue}{\SI{5}{\percent} -- \SI{10}{\percent} and \SI{5}{\percent} -- \SI{11}{\percent}, respectively,} for other conditions.

Our results \revnote{2.4}\sout{confirm}\textcolor{blue}{are in line with} the previous finding that NN-based SSI retrieval generalizes and performs poorly under clear-sky conditions. Geographically variable ground albedo between training and test sites has been proposed as a possible cause for this \cite{fracne-ssi-generalizaiton, central-europe-cnn}. Ground albedo is a plausible source of site-specific variation under clear-sky conditions, because cloud-related variability is minimal and the satellite images contain a stronger contribution from the radiation reflected off the surface around the site. However, another study considering SSI retrieval in North America did not use a geographic test-train split and still reported poor clear-sky performance. It was proposed that inadequate capture of aerosols and water vapour in selected spectral bands was a possible reason \cite{US-CNN-SSI}. We did not perform experiments with a purely temporal split, so we cannot confirm this effect with our results. In our results, the bias share of error is the highest in clear-sky conditions. This suggests that the main issue limiting clear-sky performance is likely the bias of the model. This tentatively points to the ground albedo being the cause, as opposed to the effect of aerosols and water vapour not being well-captured in the images. This is because we would expect the prevalence of water vapour and aerosols in the atmosphere to be variable, which would introduce a prediction error without bias. However, this interpretation is tentative because aerosol and water-vapour effects may also lead to a geographically heterogeneous or seasonal bias.

It is possible that in our experiment, the reason for the higher clear-sky bias and consequent poorer performance is the relatively small prevalence of clear conditions in our dataset. In particular, out of four sky condition classes, the timesteps from days labelled as ``clear'' make up only \revnote{2.4}\sout{\mbox{\SI{4}{\percent}}}\textcolor{blue}{\SI{5}{\percent} and \SI{6}{\percent}} of our \textcolor{blue}{training and test} dataset\textcolor{blue}{s, respectively (\autoref{tab:sky_condition_definitions})}. As a result, the trained model might be better calibrated for SSI estimations in cloudy conditions, producing a condition-dependent calibration bias when applied to clear-sky samples, where the input image feature distributions differ from those in cloudy regimes. This interpretation might also explain higher seed-wise variability in clear-sky conditions: different random initializations might lead to different calibrations that perform similarly in cloudy conditions, but diverge more strongly on clear-sky samples. 

%It is possible that our models learn certain internal offsets which depend on or align with input image distributions. Because most of the input images contain clouds, it is possible that these offsets would be learned in such a way as to prioritise bias-free and accurate estimation in cloudy conditions. However, in clear conditions, where the image feature distributions are different due to different optical properties of the surface compared to the cloud cover, these internal offsets would, in turn, result in a biased output. Such biases can also provide a plausible explanation for higher seed-wise performance instability in clear-sky conditions. Perhaps the learned offsets are different for each seed, but they consistently align to result in bias-free output for cloudy conditions. Under clear-sky conditions, they are, however, misaligned, resulting in SSI biases of different magnitudes.

 Another factor to consider is that clear-sky performance was mainly compared against gridded meteorological products, which rely on physics-based models. Due to the gridded nature of these products, they estimate the average solar irradiance over a comparatively large grid cell, which in the case of SARAH-3 is approximately $\SI{5.6}{\km}\times\SI{5.6}{\km}$. As such, they can be relatively more accurate for SSI estimations at a single point in a given grid cell in clear skies as compared to cloudy skies due to the lack of variability of SSI within that cell that would otherwise be caused by clouds. Because gridded meteorological products are physics-based, they might be less susceptible to bias, which in the case of a machine learning model can come from training and test input and target distribution or properties being different, and as such be comparatively further advantaged on clear-sky days. This means that gridded meteorological products can be a much stronger baseline in clear conditions compared to cloudy conditions for SSI prediction. 

Multiple limitations must be considered for the interpretation of our results. Firstly, the small geographical extent in Northern Europe limits the generality of our results and does not allow us to study the generalization ability of our models. Secondly, while the spatial resolution is a major difference between the FCI and SEVIRI imagers, other factors could have affected observed performance. These factors include the spectral resolution of the imagers, differences in imaging time alignment with SSI observations, patch sizes, and selected spectral bands that do not exactly match between imagers. As such, we cannot confirm specifically which properties of the FCI imager contributed most to the improved performance. Thirdly, \revnote{1.5, 1.8, 2.1, 2.2, 2.3, 2.4}\sout{the HYBRID model inherently has more parameters due to the additional branches, which could have improved its performance.}\textcolor{blue}{both models are trained on the same single year of data, although a longer SEVIRI record is available. The SEVIRI-FN-3Y model, which is pre-trained on three years of SEVIRI data, has less than \SI{1}{\percent} higher all-sky RMSE than the FCI-only model (\autoref{sec:three-year}), although this difference was not tested for statistical significance. As such, our main comparison shows the advantage of FCI imagery when both imagers are trained on the same period, and a longer SEVIRI training record combined with fine-tuning on the year preceding the test period might close most of this advantage in all-sky conditions.} \textcolor{blue}{Fourthly, EUMETSAT changed the calibration of the SEVIRI solar channels during our test period, and we recalibrated the affected SEVIRI images to the calibration of the training period. As such, our SEVIRI test images differ from those distributed by EUMETSAT, and CNN-based models applied to newly distributed SEVIRI data might require a similar conversion, or retraining, after each calibration change.} \textcolor{blue}{Fifthly, clear-sky timesteps make up only approximately \SI{5}{\percent} of our training data and \SI{6}{\percent} of our test data (\autoref{tab:sky_condition_definitions}), and the clear-sky RMSE of our models varies strongly between training seeds (\autoref{tab:cv-bias-share}). As such, our clear-sky results, and in particular the clear-sky differences between our models, are less certain than those for other sky conditions. We did not perform experiments that increase the weight of clear-sky samples during training.} Finally, due to limited FCI data availability, only \sout{data from the year 2025 could be used for fine-tuning and evaluation. If this year's weather patterns were significantly different from those of other years in the SEVIRI dataset, this could have biased our results with the HYBRID model to yield better performance compared to the SEVIRI-only model, which was trained on data from 5 years. Limited FCI data availability additionally necessitated a temporal overlap between test and train splits. Due to shared general weather patterns between the splits, this overlap could have inflated the performance for both models we tested.}\textcolor{blue}{approximately one year of data could be used for training and the following year for testing. A single training year might not represent the interannual variability of weather patterns, and the training outcomes of models trained on a single year were observed to be more unstable in preliminary testing. Additionally, the validation sites share the training period, so model selection does not account for the temporal shift between training and testing.}

As more FCI data becomes available, future work should focus on evaluating CNN-based SSI retrieval models with longer FCI time series and independent test years in a larger geographic domain. Additionally, further studies should prioritize explaining clear-sky performance shortfalls in detail and improving model performance in these conditions. For example, the importance of clear-sky samples could be increased during training with synthetic or repeated datapoints. We have shown that deep learning models can improve SSI estimates under cloudy skies, and next-generation geostationary imagers like MTG/FCI can further improve SSI retrieval \revnote{2.1}\sout{when combined with previous-generation data such as MSG/SEVIRI in a two-imager hybrid approach}\textcolor{blue}{compared to previous-generation imagers such as MSG/SEVIRI}. We therefore believe that a promising research avenue would be to extend the proposed architecture for SSI forecasting with a considerable lead time. This could be done, for example, by adding recurrent or attention elements over the embeddings of images produced by the branches of the SSI retrieval model. 

In summary, our results show that higher-resolution MTG/FCI imagery can improve CNN-based SSI retrieval when clouds are the dominant source of irradiance variability. At the same time, the lack of improvement under clear skies \revnote{1.3, 1.7, 2.1, 2.4}\sout{shows}\textcolor{blue}{suggests} that higher spatial resolution alone is not sufficient to solve all limitations of machine-learning-based SSI retrieval. \textcolor{blue}{Additionally, a SEVIRI-only model pre-trained on a longer SEVIRI record and fine-tuned on the year preceding the test period approached the all-sky RMSE of the FCI-only model, so the advantage of FCI imagery that we found applies to models trained on the same period.} Our main contributions are the inclusion of FCI data, the multi-resolution \sout{and multi-imager} model architecture, and robust cross-validation of results.

\section*{Declaration of generative AI and AI-assisted technologies in the manuscript preparation process} 
During the preparation of this work, the lead author used ChatGPT \textcolor{blue}{and Claude Code to assist with writing}. Additionally, the Cursor IDE \textcolor{blue}{and Claude Code} were used to assist with conceptualizing some aspects of the study methodology, data analysis, and creation of plots and tables.  After using AI tools, the authors critically reviewed and edited the content as needed and take full responsibility for the content of the published article.

\section*{CRediT authorship contribution statement} \textbf{Gordei Pribõtkin:} Conceptualization, Methodology, Software, Data curation, Formal analysis, Investigation, Resources, Validation, Visualization, Writing -- original draft, Writing -- review \& editing. 

\textbf{Piia Post:} Conceptualization, Methodology, Supervision, Project administration, Writing -- review \& editing. 

\textbf{Velle Toll:} Conceptualization, Methodology, Supervision, Project administration, Writing -- review \& editing.

\section*{Declaration of competing interest}
We declare that G.P. was supported by STACC OÜ, which sourced institutional research funding and provided computational resources. The funder had no role in the study design, data analysis, interpretation of results, or preparation of the manuscript.

\section*{Funding}
STACC OÜ supported G.P. with regular base funding for R\&D from the Ministry of Education and Research (Estonia).

\section*{Acknowledgements}
The authors gratefully acknowledge the Estonian Environment Agency for providing quality-controlled ground-based irradiance measurements used in this study. The authors also acknowledge EUMETSAT for access to Meteosat Second Generation SEVIRI and Meteosat Third Generation FCI data, and the CM SAF/EUMETSAT SARAH-3 surface solar radiation dataset used for baseline comparison.

%Bibliography
\bibliographystyle{unsrt}  
\bibliography{references}

\appendix

\section{HYBRID model}
\label{sec:hybrid-scratch}
\textcolor{blue}{\revnote{2.1, 2.3}In addition to the SEVIRI-only and FCI-only models, we train a HYBRID model, which receives both SEVIRI and FCI images. The HYBRID model is trained from randomly initialized weights on the same period, from 2024.09.24 to 2025.09.23, and shares the architecture, numerical inputs, training procedure and cross-validation folds of the SEVIRI-only and FCI-only models. The HYBRID model is trained from scratch, meaning that no weights are taken from the SEVIRI-only model and no branches are frozen, in contrast to the fine-tuned models in \autoref{sec:three-year}. The model has 945k trainable parameters.}

\begin{figure}[h!]
    \centering
    \includegraphics[width=0.5\linewidth]{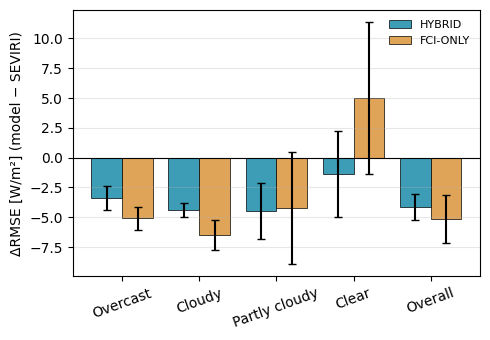}
    \caption{\revnote{2.1, 2.3}\textcolor{blue}{Paired RMSE difference (model -- SEVIRI-only) of the FCI-only (labelled FCI-ONLY) and HYBRID models by sky condition and overall. Negative values indicate lower RMSE than the SEVIRI-only model (fold-level mean $\pm$ SEM; seeds averaged within each fold).}}
    \label{fig:deltaRMSE-hybrid-scratch}
\end{figure}

\textcolor{blue}{\revnote{2.1, 2.3}The HYBRID model has not yielded an improvement over the FCI-only model. It results in lower mean RMSE than the SEVIRI-only model in all sky conditions, but higher mean RMSE than the FCI-only model in overcast, cloudy and all-sky conditions. \autoref{fig:deltaRMSE-hybrid-scratch} shows the paired RMSE difference of the FCI-only and HYBRID models relative to the SEVIRI-only model, and \autoref{tab:cv-overall-hybrid-scratch} lists the aggregated overall metrics of all three models. Compared with the SEVIRI-only model, the HYBRID model has \SI{3.4}{\watt\per\square\meter}, \SI{4.3}{\watt\per\square\meter}, \SI{4.5}{\watt\per\square\meter}, \SI{1.4}{\watt\per\square\meter} and \SI{4.1}{\watt\per\square\meter} lower RMSE in overcast, cloudy, partly cloudy, clear and all-sky conditions, respectively, against \SI{5.1}{\watt\per\square\meter}, \SI{6.5}{\watt\per\square\meter}, \SI{4.2}{\watt\per\square\meter} and \SI{5.1}{\watt\per\square\meter} lower RMSE in overcast, cloudy, partly cloudy and all-sky conditions and \SI{5.0}{\watt\per\square\meter} higher RMSE in clear conditions for the FCI-only model. We did not test the HYBRID differences for statistical significance, so they are reported as descriptive only. Additionally, the \SI{6.4}{\watt\per\square\meter} lower clear-sky RMSE of the HYBRID model compared with the FCI-only model should be read with caution, given the high seed-wise variability of clear-sky results shown in \autoref{tab:cv-bias-share}. Under partly cloudy skies, the mean RMSE of the HYBRID and FCI-only models differs by only \SI{0.2}{\watt\per\square\meter}. The overall MBE of the HYBRID model, \SI{-2.2}{\watt\per\square\meter}, is close to that of the FCI-only model, \SI{-2.5}{\watt\per\square\meter}, and smaller in magnitude than that of the SEVIRI-only model, \SI{-4.6}{\watt\per\square\meter}. A possible reason for the smaller improvement is that, when both imagers are available from the start of training, the HYBRID model might rely partly on the SEVIRI branches and make less use of the higher-resolution FCI images. However, this interpretation is tentative, as we did not analyse the contribution of the individual branches.}

\begin{table}[h!]
\centering
\color{blue}
\caption{\revnote{2.1, 2.3}Aggregated overall (all-sky) metrics of the SEVIRI-only, FCI-only and HYBRID models. Values are means over all folds and training seeds; standard deviations are not shown. RMSE, MAE and MBE are computed on all test timestamps. The skill score SS is computed against SARAH-3 on the subset of timestamps where SARAH-3 is available; positive values indicate lower RMSE than SARAH-3.}
\label{tab:cv-overall-hybrid-scratch}
\begin{tabularx}{\linewidth}{X c c c c}
\toprule
Model & RMSE [W/m$^2$] & MAE [W/m$^2$] & MBE [W/m$^2$] & SS [\%] \\
\midrule
SEVIRI-only & $85.52$ & $52.64$ & $-4.58$ & $11.25$ \\
FCI-only & $80.38$ & $49.58$ & $-2.54$ & $16.90$ \\
HYBRID & $81.38$ & $50.28$ & $-2.22$ & $15.85$ \\
\bottomrule
\end{tabularx}
\end{table}

\FloatBarrier
\newpage
\section{Performance metrics by fold}
\begin{table}[h!]
\centering
\caption{\revnote{1.5, 2.2}Per-fold RMSE for each held-out test site, by sky condition \textcolor{blue}{and overall}, for the SEVIRI-only\textcolor{blue}{, FCI-only and HYBRID (see \autoref{sec:hybrid-scratch})} models and SARAH-3. Metrics are computed on the subset of timestamps where SARAH-3 is available, so all \textcolor{blue}{four} sources are directly comparable. \textcolor{blue}{Overall values are computed on all these timestamps pooled together, not averaged over the sky conditions.} Training seeds are averaged within each fold.}
\label{tab:cv-rmse-perfold-by-sky}
\begin{tabularx}{\linewidth}{l p{3cm} X X X X X}
\toprule
Test site & Model & Overcast & Cloudy & Partly cloudy & Clear & \textcolor{blue}{Overall} \\
\midrule
\multirow{4}{*}{Haapsalu} & SEVIRI-only & \textcolor{blue}{$49.94$} & \textcolor{blue}{$101.83$} & \textcolor{blue}{$74.54$} & \textcolor{blue}{$27.14$} & \textcolor{blue}{$79.02$} \\
 & \textcolor{blue}{FCI-only} & \textcolor{blue}{$45.51$} & \textcolor{blue}{$95.96$} & \textcolor{blue}{$74.38$} & \textcolor{blue}{$36.50$} & \textcolor{blue}{$75.61$} \\
 & \textcolor{blue}{HYBRID} & \textcolor{blue}{$49.86$} & \textcolor{blue}{$99.37$} & \textcolor{blue}{$73.33$} & \textcolor{blue}{$26.79$} & \textcolor{blue}{$77.47$} \\
 & SARAH-3 & \textcolor{blue}{$74.44$} & \textcolor{blue}{$126.39$} & \textcolor{blue}{$86.52$} & \textcolor{blue}{$26.11$} & \textcolor{blue}{$98.88$} \\
\midrule
\multirow{4}{*}{Tiirikoja} & SEVIRI-only & \textcolor{blue}{$61.88$} & \textcolor{blue}{$105.18$} & \textcolor{blue}{$90.61$} & \textcolor{blue}{$40.05$} & \textcolor{blue}{$88.72$} \\
 & \textcolor{blue}{FCI-only} & \textcolor{blue}{$60.34$} & \textcolor{blue}{$103.44$} & \textcolor{blue}{$102.75$} & \textcolor{blue}{$74.05$} & \textcolor{blue}{$91.94$} \\
 & \textcolor{blue}{HYBRID} & \textcolor{blue}{$57.20$} & \textcolor{blue}{$100.72$} & \textcolor{blue}{$89.68$} & \textcolor{blue}{$45.58$} & \textcolor{blue}{$85.44$} \\
 & SARAH-3 & \textcolor{blue}{$79.61$} & \textcolor{blue}{$122.12$} & \textcolor{blue}{$95.45$} & \textcolor{blue}{$31.39$} & \textcolor{blue}{$102.03$} \\
\midrule
\multirow{4}{*}{Tõravere} & SEVIRI-only & \textcolor{blue}{$59.41$} & \textcolor{blue}{$104.86$} & \textcolor{blue}{$85.91$} & \textcolor{blue}{$32.92$} & \textcolor{blue}{$85.43$} \\
 & \textcolor{blue}{FCI-only} & \textcolor{blue}{$57.87$} & \textcolor{blue}{$102.17$} & \textcolor{blue}{$86.06$} & \textcolor{blue}{$43.25$} & \textcolor{blue}{$84.14$} \\
 & \textcolor{blue}{HYBRID} & \textcolor{blue}{$57.33$} & \textcolor{blue}{$101.05$} & \textcolor{blue}{$84.69$} & \textcolor{blue}{$42.81$} & \textcolor{blue}{$83.15$} \\
 & SARAH-3 & \textcolor{blue}{$77.68$} & \textcolor{blue}{$127.55$} & \textcolor{blue}{$94.69$} & \textcolor{blue}{$29.87$} & \textcolor{blue}{$102.52$} \\
\midrule
\multirow{4}{*}{Pärnu} & SEVIRI-only & \textcolor{blue}{$54.89$} & \textcolor{blue}{$107.34$} & \textcolor{blue}{$76.69$} & \textcolor{blue}{$36.97$} & \textcolor{blue}{$86.11$} \\
 & \textcolor{blue}{FCI-only} & \textcolor{blue}{$47.29$} & \textcolor{blue}{$93.93$} & \textcolor{blue}{$67.01$} & \textcolor{blue}{$34.02$} & \textcolor{blue}{$75.16$} \\
 & \textcolor{blue}{HYBRID} & \textcolor{blue}{$55.48$} & \textcolor{blue}{$102.15$} & \textcolor{blue}{$69.79$} & \textcolor{blue}{$34.60$} & \textcolor{blue}{$81.89$} \\
 & SARAH-3 & \textcolor{blue}{$71.59$} & \textcolor{blue}{$126.33$} & \textcolor{blue}{$79.14$} & \textcolor{blue}{$32.03$} & \textcolor{blue}{$100.06$} \\
\midrule
\multirow{4}{*}{Roomassaare} & SEVIRI-only & \textcolor{blue}{$59.33$} & \textcolor{blue}{$96.82$} & \textcolor{blue}{$75.48$} & \textcolor{blue}{$40.23$} & \textcolor{blue}{$78.85$} \\
 & \textcolor{blue}{FCI-only} & \textcolor{blue}{$50.20$} & \textcolor{blue}{$90.45$} & \textcolor{blue}{$75.96$} & \textcolor{blue}{$42.75$} & \textcolor{blue}{$74.57$} \\
 & \textcolor{blue}{HYBRID} & \textcolor{blue}{$54.17$} & \textcolor{blue}{$92.56$} & \textcolor{blue}{$73.69$} & \textcolor{blue}{$34.76$} & \textcolor{blue}{$75.13$} \\
 & SARAH-3 & \textcolor{blue}{$71.29$} & \textcolor{blue}{$115.53$} & \textcolor{blue}{$81.68$} & \textcolor{blue}{$33.15$} & \textcolor{blue}{$91.21$} \\
\midrule
\multirow{4}{*}{Vilsandi} & SEVIRI-only & \textcolor{blue}{$61.02$} & \textcolor{blue}{$102.39$} & \textcolor{blue}{$103.16$} & \textcolor{blue}{$89.86$} & \textcolor{blue}{$93.82$} \\
 & \textcolor{blue}{FCI-only} & \textcolor{blue}{$55.62$} & \textcolor{blue}{$92.37$} & \textcolor{blue}{$70.84$} & \textcolor{blue}{$62.27$} & \textcolor{blue}{$78.33$} \\
 & \textcolor{blue}{HYBRID} & \textcolor{blue}{$52.89$} & \textcolor{blue}{$95.03$} & \textcolor{blue}{$83.07$} & \textcolor{blue}{$66.25$} & \textcolor{blue}{$81.81$} \\
 & SARAH-3 & \textcolor{blue}{$70.54$} & \textcolor{blue}{$102.62$} & \textcolor{blue}{$59.34$} & \textcolor{blue}{$39.52$} & \textcolor{blue}{$84.28$} \\
\midrule
\multirow{4}{*}{Tallinn} & SEVIRI-only & \textcolor{blue}{$57.00$} & \textcolor{blue}{$111.29$} & \textcolor{blue}{$86.24$} & \textcolor{blue}{$48.46$} & \textcolor{blue}{$90.17$} \\
 & \textcolor{blue}{FCI-only} & \textcolor{blue}{$53.91$} & \textcolor{blue}{$101.95$} & \textcolor{blue}{$79.56$} & \textcolor{blue}{$42.58$} & \textcolor{blue}{$83.01$} \\
 & \textcolor{blue}{HYBRID} & \textcolor{blue}{$54.87$} & \textcolor{blue}{$104.70$} & \textcolor{blue}{$85.26$} & \textcolor{blue}{$52.50$} & \textcolor{blue}{$86.25$} \\
 & SARAH-3 & \textcolor{blue}{$73.51$} & \textcolor{blue}{$129.35$} & \textcolor{blue}{$84.79$} & \textcolor{blue}{$32.32$} & \textcolor{blue}{$102.16$} \\
\midrule
\multirow{4}{*}{Narva} & SEVIRI-only & \textcolor{blue}{$70.19$} & \textcolor{blue}{$102.63$} & \textcolor{blue}{$80.14$} & \textcolor{blue}{$39.61$} & \textcolor{blue}{$85.30$} \\
 & \textcolor{blue}{FCI-only} & \textcolor{blue}{$63.17$} & \textcolor{blue}{$98.65$} & \textcolor{blue}{$82.39$} & \textcolor{blue}{$57.47$} & \textcolor{blue}{$82.77$} \\
 & \textcolor{blue}{HYBRID} & \textcolor{blue}{$66.20$} & \textcolor{blue}{$99.79$} & \textcolor{blue}{$75.75$} & \textcolor{blue}{$38.13$} & \textcolor{blue}{$81.98$} \\
 & SARAH-3 & \textcolor{blue}{$83.23$} & \textcolor{blue}{$120.08$} & \textcolor{blue}{$79.95$} & \textcolor{blue}{$31.04$} & \textcolor{blue}{$97.21$} \\
\bottomrule
\end{tabularx}
\end{table}

\begin{table}[h!]
\centering
\caption{\revnote{1.5, 2.2}Per-fold MBE [W/m$^2$] for each held-out test site, by sky condition \textcolor{blue}{and overall}, for the SEVIRI-only\textcolor{blue}{, FCI-only and HYBRID (see \autoref{sec:hybrid-scratch})} models and SARAH-3. Metrics are computed on the subset of timestamps where SARAH-3 is available, so all \textcolor{blue}{four} sources are directly comparable. \textcolor{blue}{Overall values are computed on all these timestamps pooled together, not averaged over the sky conditions.} Training seeds are averaged within
each fold.}
\label{tab:cv-mbe-perfold-by-sky}
\begin{tabularx}{\linewidth}{l p{3cm} X X X X X}
\toprule
Test site & Model & Overcast & Cloudy & Partly cloudy & Clear & \textcolor{blue}{Overall} \\
\midrule
\multirow{4}{*}{Haapsalu} & SEVIRI-only & \textcolor{blue}{$-1.02$} & \textcolor{blue}{$-5.30$} & \textcolor{blue}{$-6.94$} & \textcolor{blue}{$-6.52$} & \textcolor{blue}{$-4.62$} \\
 & \textcolor{blue}{FCI-only} & \textcolor{blue}{$7.05$} & \textcolor{blue}{$-6.37$} & \textcolor{blue}{$-16.37$} & \textcolor{blue}{$-23.37$} & \textcolor{blue}{$-6.41$} \\
 & \textcolor{blue}{HYBRID} & \textcolor{blue}{$0.67$} & \textcolor{blue}{$-3.25$} & \textcolor{blue}{$-2.32$} & \textcolor{blue}{$-5.14$} & \textcolor{blue}{$-2.04$} \\
 & SARAH-3 & \textcolor{blue}{$12.80$} & \textcolor{blue}{$-5.32$} & \textcolor{blue}{$-9.87$} & \textcolor{blue}{$-6.49$} & \textcolor{blue}{$-1.53$} \\
\midrule
\multirow{4}{*}{Tiirikoja} & SEVIRI-only & \textcolor{blue}{$-5.55$} & \textcolor{blue}{$-3.99$} & \textcolor{blue}{$-18.99$} & \textcolor{blue}{$-22.04$} & \textcolor{blue}{$-8.88$} \\
 & \textcolor{blue}{FCI-only} & \textcolor{blue}{$-1.01$} & \textcolor{blue}{$-8.25$} & \textcolor{blue}{$-41.91$} & \textcolor{blue}{$-52.42$} & \textcolor{blue}{$-16.31$} \\
 & \textcolor{blue}{HYBRID} & \textcolor{blue}{$7.95$} & \textcolor{blue}{$4.94$} & \textcolor{blue}{$-14.15$} & \textcolor{blue}{$-23.38$} & \textcolor{blue}{$-0.10$} \\
 & SARAH-3 & \textcolor{blue}{$6.79$} & \textcolor{blue}{$-14.06$} & \textcolor{blue}{$-19.67$} & \textcolor{blue}{$-10.58$} & \textcolor{blue}{$-9.26$} \\
\midrule
\multirow{4}{*}{Tõravere} & SEVIRI-only & \textcolor{blue}{$7.13$} & \textcolor{blue}{$-8.81$} & \textcolor{blue}{$-17.61$} & \textcolor{blue}{$-24.06$} & \textcolor{blue}{$-7.10$} \\
 & \textcolor{blue}{FCI-only} & \textcolor{blue}{$-1.55$} & \textcolor{blue}{$-11.30$} & \textcolor{blue}{$-18.44$} & \textcolor{blue}{$-31.81$} & \textcolor{blue}{$-11.34$} \\
 & \textcolor{blue}{HYBRID} & \textcolor{blue}{$9.55$} & \textcolor{blue}{$-7.15$} & \textcolor{blue}{$-18.85$} & \textcolor{blue}{$-33.21$} & \textcolor{blue}{$-6.56$} \\
 & SARAH-3 & \textcolor{blue}{$12.75$} & \textcolor{blue}{$-7.99$} & \textcolor{blue}{$-10.45$} & \textcolor{blue}{$-4.77$} & \textcolor{blue}{$-2.20$} \\
\midrule
\multirow{4}{*}{Pärnu} & SEVIRI-only & \textcolor{blue}{$12.05$} & \textcolor{blue}{$-0.59$} & \textcolor{blue}{$-7.35$} & \textcolor{blue}{$-5.34$} & \textcolor{blue}{$1.02$} \\
 & \textcolor{blue}{FCI-only} & \textcolor{blue}{$7.05$} & \textcolor{blue}{$4.30$} & \textcolor{blue}{$7.92$} & \textcolor{blue}{$5.61$} & \textcolor{blue}{$5.97$} \\
 & \textcolor{blue}{HYBRID} & \textcolor{blue}{$15.42$} & \textcolor{blue}{$5.55$} & \textcolor{blue}{$-4.46$} & \textcolor{blue}{$-6.63$} & \textcolor{blue}{$5.29$} \\
 & SARAH-3 & \textcolor{blue}{$13.04$} & \textcolor{blue}{$-3.92$} & \textcolor{blue}{$-5.65$} & \textcolor{blue}{$3.86$} & \textcolor{blue}{$0.63$} \\
\midrule
\multirow{4}{*}{Roomassaare} & SEVIRI-only & \textcolor{blue}{$11.50$} & \textcolor{blue}{$1.80$} & \textcolor{blue}{$-7.02$} & \textcolor{blue}{$0.88$} & \textcolor{blue}{$1.49$} \\
 & \textcolor{blue}{FCI-only} & \textcolor{blue}{$-4.55$} & \textcolor{blue}{$-10.01$} & \textcolor{blue}{$-1.13$} & \textcolor{blue}{$15.31$} & \textcolor{blue}{$-3.87$} \\
 & \textcolor{blue}{HYBRID} & \textcolor{blue}{$3.29$} & \textcolor{blue}{$-5.63$} & \textcolor{blue}{$-6.81$} & \textcolor{blue}{$1.06$} & \textcolor{blue}{$-3.23$} \\
 & SARAH-3 & \textcolor{blue}{$7.16$} & \textcolor{blue}{$-13.89$} & \textcolor{blue}{$-18.21$} & \textcolor{blue}{$-8.88$} & \textcolor{blue}{$-9.62$} \\
\midrule
\multirow{4}{*}{Vilsandi} & SEVIRI-only & \textcolor{blue}{$19.31$} & \textcolor{blue}{$-6.07$} & \textcolor{blue}{$-57.06$} & \textcolor{blue}{$-51.84$} & \textcolor{blue}{$-12.31$} \\
 & \textcolor{blue}{FCI-only} & \textcolor{blue}{$15.30$} & \textcolor{blue}{$23.08$} & \textcolor{blue}{$-4.49$} & \textcolor{blue}{$5.82$} & \textcolor{blue}{$14.44$} \\
 & \textcolor{blue}{HYBRID} & \textcolor{blue}{$13.21$} & \textcolor{blue}{$2.82$} & \textcolor{blue}{$-49.60$} & \textcolor{blue}{$-45.79$} & \textcolor{blue}{$-7.87$} \\
 & SARAH-3 & \textcolor{blue}{$12.55$} & \textcolor{blue}{$9.72$} & \textcolor{blue}{$-0.92$} & \textcolor{blue}{$8.59$} & \textcolor{blue}{$8.27$} \\
\midrule
\multirow{4}{*}{Tallinn} & SEVIRI-only & \textcolor{blue}{$2.66$} & \textcolor{blue}{$-20.75$} & \textcolor{blue}{$-23.77$} & \textcolor{blue}{$-24.29$} & \textcolor{blue}{$-14.69$} \\
 & \textcolor{blue}{FCI-only} & \textcolor{blue}{$-9.43$} & \textcolor{blue}{$-21.29$} & \textcolor{blue}{$-20.50$} & \textcolor{blue}{$-29.17$} & \textcolor{blue}{$-17.72$} \\
 & \textcolor{blue}{HYBRID} & \textcolor{blue}{$-3.52$} & \textcolor{blue}{$-19.77$} & \textcolor{blue}{$-34.56$} & \textcolor{blue}{$-42.95$} & \textcolor{blue}{$-19.45$} \\
 & SARAH-3 & \textcolor{blue}{$8.39$} & \textcolor{blue}{$-11.12$} & \textcolor{blue}{$-10.14$} & \textcolor{blue}{$-12.97$} & \textcolor{blue}{$-5.09$} \\
\midrule
\multirow{4}{*}{Narva} & SEVIRI-only & \textcolor{blue}{$7.88$} & \textcolor{blue}{$3.36$} & \textcolor{blue}{$6.70$} & \textcolor{blue}{$16.51$} & \textcolor{blue}{$6.44$} \\
 & \textcolor{blue}{FCI-only} & \textcolor{blue}{$2.70$} & \textcolor{blue}{$8.68$} & \textcolor{blue}{$28.52$} & \textcolor{blue}{$42.60$} & \textcolor{blue}{$13.12$} \\
 & \textcolor{blue}{HYBRID} & \textcolor{blue}{$15.34$} & \textcolor{blue}{$13.95$} & \textcolor{blue}{$12.14$} & \textcolor{blue}{$17.15$} & \textcolor{blue}{$14.29$} \\
 & SARAH-3 & \textcolor{blue}{$9.88$} & \textcolor{blue}{$-7.75$} & \textcolor{blue}{$-2.57$} & \textcolor{blue}{$3.97$} & \textcolor{blue}{$-0.20$} \\
\bottomrule
\end{tabularx}
\end{table}

\FloatBarrier

\section{Models trained on a longer SEVIRI record}
\label{sec:three-year}

\begin{table}[h!]
\centering
\color{blue}
\caption{\revnote{2.1}RMSE [W/m$^2$] of the one-year models (SEVIRI-only, FCI-only and HYBRID, trained on 2024.09.24--2025.09.23) and the three-year models (SEVIRI-only-3Y, SEVIRI-FN-3Y and HYBRID-FN-3Y, with SEVIRI pre-training on 2022.09.24--2025.09.23), by sky condition and overall. The suffix FN denotes fine-tuning from the SEVIRI-only-3Y model on 2024.09.24--2025.09.23. Values are means over all folds and training seeds; standard deviations are not shown. RMSE is computed on all test timestamps from 2025.09.24 to 2026.09.12, without restriction to SARAH-3 availability.}
\label{tab:cv-rmse-3y}
\begin{tabularx}{\linewidth}{l X c c c c c}
\toprule
Training data & Model & Overcast & Cloudy & Partly cloudy & Clear & Overall \\
\midrule
\multirow{3}{*}{One year} & SEVIRI-only & $59.58$ & $103.41$ & $83.52$ & $44.33$ & $85.52$ \\
 & FCI-only & $54.48$ & $96.94$ & $79.30$ & $49.34$ & $80.38$ \\
 & HYBRID & $56.20$ & $99.06$ & $79.05$ & $42.95$ & $81.38$ \\
\midrule
\multirow{3}{*}{Three years} & SEVIRI-only-3Y & $56.01$ & $103.52$ & $82.69$ & $42.90$ & $84.57$ \\
 & SEVIRI-FN-3Y & $50.72$ & $98.68$ & $82.16$ & $45.36$ & $81.02$ \\
 & HYBRID-FN-3Y & $51.06$ & $99.24$ & $82.92$ & $45.32$ & $81.61$ \\
\bottomrule
\end{tabularx}
\end{table}

\textcolor{blue}{\revnote{2.1}The SEVIRI-only and FCI-only models in the main text are trained on a single year, although a longer SEVIRI record is available. To test whether the lower RMSE of the FCI-only model holds against a SEVIRI-only model trained on multiple years, we additionally train three models, to which we refer as three-year models and which are marked with the suffix 3Y. The SEVIRI-only-3Y model receives only SEVIRI images and is trained from randomly initialized weights on three years of data, from 2022.09.24 to 2025.09.23. To obtain the HYBRID-FN-3Y model, we fix the SEVIRI branch parameters of the SEVIRI-only-3Y model and train the FCI branches and the head on the period from 2024.09.24 to 2025.09.23, where FCI is available. The previously-untrained FCI branch parameters use a base learning rate of \num{1e-3},  while the head parameters train with a base learning rate of \num{1e-4}. To obtain the SEVIRI-FN-3Y model, we fine-tune all weights of the SEVIRI-only-3Y model on SEVIRI data from the same period, without FCI images. Here, the head and SEVIRI branches train with a base learning rate of \num{1e-4}. The suffix FN denotes fine-tuning from the SEVIRI-only-3Y model, in contrast to training from scratch (\autoref{sec:hybrid-scratch}). We chose to include the SEVIRI-FN-3Y model because the HYBRID-FN-3Y model receives additional training on the year preceding the test period, and the SEVIRI-FN-3Y model receives the same additional training without FCI images. This allows us to separate the effect of the FCI images from the effect of the additional training. The three-year models share the architecture, numerical inputs, cross-validation folds and test period, from 2025.09.24 to 2026.09.12, of the one-year models, but are trained with four seeds per fold instead of five.}

\textcolor{blue}{\revnote{2.1}A longer SEVIRI training record lowers the RMSE of the SEVIRI-only model mainly under overcast and clear skies, and the FCI-only model retains the lowest RMSE only in cloudy and all-sky conditions. \autoref{tab:cv-rmse-3y} lists the RMSE of the one-year and three-year models by sky condition and overall. Overall, the SEVIRI-only-3Y model has only \SI{1.0}{\watt\per\square\meter} lower RMSE than the SEVIRI-only model, with little change in cloudy and partly cloudy conditions. Among the three-year models, the SEVIRI-FN-3Y model has the lowest RMSE in all conditions except clear skies, with \SI{3.6}{\watt\per\square\meter} lower overall RMSE than the SEVIRI-only-3Y model. The HYBRID-FN-3Y model has \SI{3.0}{\watt\per\square\meter} lower overall RMSE than the SEVIRI-only-3Y model, but \SI{0.6}{\watt\per\square\meter} higher overall RMSE than the SEVIRI-FN-3Y model, and it is higher in every condition except clear skies, where the two models differ by less than \SI{0.1}{\watt\per\square\meter}. The FCI-only model in turn has \SI{0.6}{\watt\per\square\meter} lower overall RMSE than the SEVIRI-FN-3Y model, with \SI{1.7}{\watt\per\square\meter} and \SI{2.9}{\watt\per\square\meter} lower RMSE in cloudy and partly cloudy conditions, respectively, but \SI{3.8}{\watt\per\square\meter} higher RMSE under overcast skies. Under clear skies, the RMSE of all six models lies between \SI{42.9}{\watt\per\square\meter} and \SI{49.3}{\watt\per\square\meter}, and given the high seed-wise variability of clear-sky results (\autoref{tab:cv-bias-share}), the clear-sky differences should be read with caution. We did not test the differences in this section for statistical significance, so they are reported as descriptive only. This suggests that a SEVIRI-only model trained on a longer record and adapted to the year preceding the test period might approach the overall performance of the FCI-only model, and that adding FCI branches to the SEVIRI-only-3Y model does not lower the RMSE further. However, this interpretation is tentative, as the differences were not tested and the improvement of the SEVIRI-FN-3Y model could stem either from adaptation to the more recent year or from the longer total training.}

\end{document}